\documentclass[10pt, a4paper, oneside]{article}
\usepackage[hidelinks]{hyperref}
\usepackage{jucs2e}
\usepackage{fancyhdr}
\usepackage{setspace}
\usepackage{graphicx}
\usepackage{url}
\usepackage{ulem}
\usepackage{mathtools}
\usepackage{scalerel}
\usepackage{setspace}
\usepackage[strict]{changepage}
\usepackage{caption}
\usepackage[letterspace=-50]{microtype}
\usepackage{afterpage}
\usepackage{ragged2e}
\usepackage{subfigure}
\usepackage[T1]{fontenc}
\usepackage[utf8]{inputenc}

\usepackage{newtxtext,newtxmath}

\usepackage{multirow}
\usepackage{rotating,tabularx}

\usepackage{url}
\usepackage{CJK}
\usepackage{listings}
\usepackage{graphicx}
\usepackage{graphicx}
\usepackage{comment}
\usepackage[utf8]{inputenc}
\usepackage{amsmath}
\usepackage{color}
\usepackage{url}
\usepackage{textcomp}

\usepackage{amssymb}
\usepackage{float}
\usepackage{longtable}
\usepackage{lscape}
\usepackage{listings}
\usepackage{tcolorbox}
\usepackage{wasysym}
\usepackage{tcolorbox}
\usepackage{multirow}

\usepackage{iftex}

\ifPDFTeX
  \usepackage[T1]{fontenc}
  \usepackage[utf8]{inputenc}
  \usepackage{newtxtext,newtxmath}
\else
  \usepackage{fontspec}
\fi

\usepackage{titlesec}

\titleformat*{\section}{\Large\bfseries}
\titleformat*{\subsection}{\normalsize\bfseries}

\graphicspath{{./figures/}}

\usepackage[textwidth=8cm, margin=0cm, left=4.6cm, right=4.2cm, top=3.9cm, bottom=6.8cm, a4paper, headheight=0.5cm, headsep=0.5cm]{geometry}
\usepackage[format=plain, labelfont=it, textfont=it, justification=centering]{caption}
\usepackage{breakcites}
\usepackage{microtype}
 
\apptocmd{\frame}{}{\justifying}{}

\newcommand\jucs{{Journal of Universal Computer Science}}
\newcommand\jucsvol{vol. 32, no. 11 (2026)}
\newcommand\jucspages{1613-1650}
\newcommand\jucssubmitted{14/8/2025}
\newcommand\jucsaccepted{5/2/2026}
\newcommand\jucsappeared{28/10/2026}
\newcommand\jucslicence{ CC BY 4.0}
\newcommand\startingPage{1613}
\newcommand\paperauthor{{Welter, C., Farias, K.: }}

\newcommand\papertitle{Model of Things (MoT): A Model-Driven Approach to Streamline Cloud-of-Things Application Development}

\header{\paperauthor \papertitle}

\begin{document}

\title{{\fontsize{14pt}{14pt}\selectfont{\vspace*{-3mm}\papertitle\vspace*{-1mm}}}}

\author{{\bfseries\fontsize{10pt}{10pt}\selectfont{Cristiano Welter}} \\
   {\fontsize{9pt}{12pt}\selectfont{(PPGCA, University of Vale do Rio dos Sinos, S\~{a}o Leopoldo, RS, Brazil\\
   cristianowelter@gmail.com)}}
   \and
   {\bfseries\fontsize{10pt}{10pt}\selectfont{Kleinner Farias}}\\
   {\fontsize{9pt}{12pt}\selectfont{(PPGCA, University of Vale do Rio dos Sinos, S\~{a}o Leopoldo, RS, Brazil \\
   \orcid{0000-0003-1891-3580}, 
    kleinnerfarias@unisinos.br)}}
}

\label{first}
\maketitle

{\fontfamily{ptm}\selectfont
\begin{abstract}
{\fontsize{9pt}{9pt}\selectfont{\vspace*{-2mm}
The convergence of Cloud Computing and the Internet of Things (IoT) underpins modern, scalable, and data-intensive systems. Cloud-of-Things (CoT) applications are now pervasive across healthcare, smart cities, and industrial automation. Nevertheless, their development remains fragmented and technically demanding. Current CoT development approaches require substantial expertise in heterogeneous platforms, protocols, and deployment technologies. They offer limited model-driven support, weak automation, and poor alignment with low-code principles. As a result, developing CoT applications remains effort-intensive and error-prone. This study aims to simplify CoT application development by introducing the Model of Things (MoT). MoT integrates UML-based modeling with low-code execution concepts to reduce technical barriers while preserving automation and deployment feasibility. We designed a custom UML profile comprising nine CoT-specific stereotypes and an automated transformation pipeline. MoT was evaluated using one end-to-end case study and a Technology Acceptance Model (TAM) questionnaire with 12 participants with software modeling experience. In the evaluated case study, MoT generated executable application components from UML models, while users supplied the required configuration parameters and deployment settings. TAM results indicate strong acceptance: all 12 participants rated MoT positively for usefulness and ease of use, and qualitative feedback emphasized reduced complexity and faster development. MoT demonstrates that combining UML modeling with low-code execution is feasible and valuable for CoT systems. By enabling executable models and lowering entry barriers, MoT advances practical, model-driven CoT engineering and opens avenues for broader industrial adoption.}}
\end{abstract}}

{\fontfamily{ptm}\selectfont
\begin{keywords}
{\fontsize{9pt}{9pt}\selectfont{Model of Things, UML, Low-code, Model-based Development}}
\end{keywords}}

{\fontfamily{ptm}\selectfont
\begin{category}
{\fontsize{9pt}{9pt}\selectfont{
H.3.1, H.3.2, H.3.3, H.3.7, H.5.1}}
\end{category}}

 {\fontfamily{ptm}\selectfont
  \begin{doi}
 {\fontsize{9pt}{9pt}\selectfont{
 10.3897/jucs.168743}}
 \end{doi}}

\section{Introduction}

The \textit{Cloud of Things} (CoT) paradigm emerges from the convergence of Cloud Computing and the Internet of Things (IoT), enabling scalable processing, storage, and orchestration of data generated by interconnected devices \cite{suhailam2024cloud,aazam2016cloud}. CoT systems increasingly support critical domains such as healthcare monitoring, smart cities, industrial automation, and intelligent environments \cite{alhaidari2023cloud,yadav2024concerns}. As IoT deployments scale and become more heterogeneous, cloud infrastructure becomes essential for handling computation-intensive tasks, real-time analytics, and elastic resource management. Despite their importance, the CoT applications remain difficult to engineer, integrate, and evolve due to the diversity of devices, communication protocols, cloud services, and deployment configurations \cite{gubbi2013internet}. The demand for rapid development cycles and continuous evolution amplifies these challenges. Consequently, there is a strong need for approaches that raise the level of abstraction, promote automation, and reduce the technical expertise required to build and deploy CoT systems. Model-driven development and low-code paradigms have been identified as promising means to address these needs by enabling developers to focus on system intent rather than platform-specific implementation details \cite{alhaidari2023cloud}. However, their integration within CoT engineering remains limited and fragmented.

Although previous studies have explored formal modeling, model-driven engineering, and deployment automation for IoT and CoT systems \cite{rw01-hellal2024formal,rw05-belguidoum2022mdmsd4iot,rw06-de2020approach}, essential gaps \textit{persist}. Formal approaches often emphasize verification and correctness but offer limited accessibility for practitioners and non-expert developers \cite{rw01-hellal2024formal}. Conversely, deployment- and automation-oriented solutions prioritize execution efficiency while relying on low-level configuration artifacts, such as scripts or declarative descriptors, that bypass high-level modeling abstractions \cite{rw09-hassan2021toward}. Low-code platforms reduce development effort but frequently lack explicit model-based foundations, making reasoning, evolution, and consistency management difficult \cite{rw13-Panayiotou2025}. Moreover, existing UML-based profiles and domain-specific languages for IoT often stop at code generation or simulation, without establishing a direct, executable link between modeling constructs and cloud-ready deployment artifacts \cite{rw03-fazel2023devs,rw07-fortas2022application}. As discussed in Section~\ref{sec:trabalhos-relacionados}, the current literature still lacks approaches that simultaneously integrate UML modeling, low-code execution, and automated deployment in a coherent, tool-supported pipeline. This gap limits the practical adoption of model-driven techniques in real-world CoT development.

To address these limitations, this article introduces the \textit{Model of Things} (MoT), a tool-supported, model-based approach designed to streamline the development of Cloud-of-Things applications. MoT aims to reduce technical barriers while preserving automation and executability by combining UML-based modeling with low-code execution principles. MoT operationalizes this objective through three core elements: a custom UML profile that captures CoT-specific concerns, an automated transformation pipeline that maps models to executable artifacts, and a deployment infrastructure that supports cloud-based execution. To assess the approach's feasibility and practical value, we used a qualitative research method aligned with exploratory and engineering-oriented studies \cite{wohlin2012experimentation}. We conducted an end-to-end case study demonstrating model-to-execution transformation and used a Technology Acceptance Model (TAM) questionnaire to capture user perceptions of usefulness and ease of use \cite{davis1989perceived,marangunic2015technology}. This method allows us to assess whether MoT, as designed, is understandable, usable, and meaningful to developers with experience in software modeling and CoT-related technologies.

Our initial results indicate that MoT successfully enables the transformation of UML models into executable CoT applications without requiring manual coding or low-level deployment configuration. In our case study, the proposed toolchain generated and deployed the complete application, demonstrating end-to-end executability. Our TAM-based evaluation involved 12 participants with backgrounds in software modeling and development. All participants reported positive perceptions regarding MoT’s usefulness and ease of use, indicating strong acceptance of the approach. Participants perceived MoT as potentially useful for reducing development complexity and delivery effort; however, these benefits were not objectively measured. Qualitative feedback emphasized reduced perceived complexity, more precise separation of concerns, and faster development compared to traditional CoT workflows. The novelty of these results lies not in algorithmic optimization but in demonstrating that executable CoT systems can be derived directly from UML models enriched with low-code semantics, unlike existing approaches that either remain conceptual or focus solely on automation scripts. MoT bridges modeling abstraction and practical deployment, showing that model-driven and low-code paradigms can coexist in a coherent engineering solution.

This work makes three main contributions. First, it introduces MoT, an engineering-focused, model-based approach that integrates a custom UML profile with automated, low-code transformations tailored to developing CoT applications. This contribution benefits researchers and practitioners seeking executable model-driven solutions beyond conceptual modeling. Second, it presents a UML profile extension that encapsulates IoT and cloud concerns as first-class modeling elements, enabling consistent, automated transformations while reducing the expertise required to build CoT systems. This contribution directly supports software developers and system architects by improving clarity, reuse, and consistency. Third, the study provides empirical evidence through a case study and user assessment, offering qualitative insights into feasibility, perceived usefulness, and ease of integration. These results benefit the broader research community by clarifying the practical limits and potential of low-code, model-driven CoT development. These contributions can lay a foundation for future work on scalability, performance evaluation, and cross-platform generalization, while delivering an immediately usable, executable approach for CoT engineering.

The remainder of this article is organized as follows. Section~\ref{sec:fundamentacao-teorica} explains the main concepts used in our study. Section~\ref{sec:trabalhos-relacionados} contrasts our study with the current literature. Section~\ref{sec:mot} presents the proposed approach. Section~\ref{sec_evalution} outlines how we evaluated the proposed approach. Section~\ref{sec:conclusao} presents some conclusions and future directions.

\section{Background}
\label{sec:fundamentacao-teorica}

This section introduces two fundamental concepts: Section \ref{subsec-cloud-of-things} provides an overview of Cloud of Things (CoT), while Section \ref{subsec-node-red} discusses Node-Red and its role in facilitating IoT application development.

\subsection{Cloud of Things}
\label{subsec-cloud-of-things}

The Cloud of Things (CoT) is a paradigm integrating the Internet of Things (IoT) with cloud computing, enabling scalable, efficient management of data and resources from diverse IoT devices \cite{alhaidari2023cloud,yadav2024concerns,aazam2016cloud}. By connecting resource-constrained devices, such as sensors, to cloud infrastructure, CoT enables the processing, analysis, and storage of large volumes of data generated by IoT. This integration provides a flexible, resilient environment that enhances the performance, reliability, and adaptability of IoT applications, transforming raw data into valuable, actionable services.

Investigating CoT in model-based software development is essential for three reasons. First, model-driven approaches can streamline the development and deployment of IoT applications on cloud platforms. CoT applications can be deployed and scaled automatically by abstracting complex cloud configurations into high-level models. This reduces human error and optimizes resource use. Second, model-based development enables faster adaptation to changing requirements, a critical need for IoT systems that often integrate diverse, evolving device types and data streams. In this way, developers can quickly modify and redeploy CoT applications. For example, this can help them stay compatible with new devices and standards. Third, UML models can help to improve consistency across CoT system components, simplifying maintenance and updates. This approach fosters maintainable architectures in which changes to IoT devices or cloud services are quickly reflected in a unified, model-driven design, improving system longevity and reducing costs.

\begin{figure}[!h]
\centering
\includegraphics[width=0.9\textwidth]{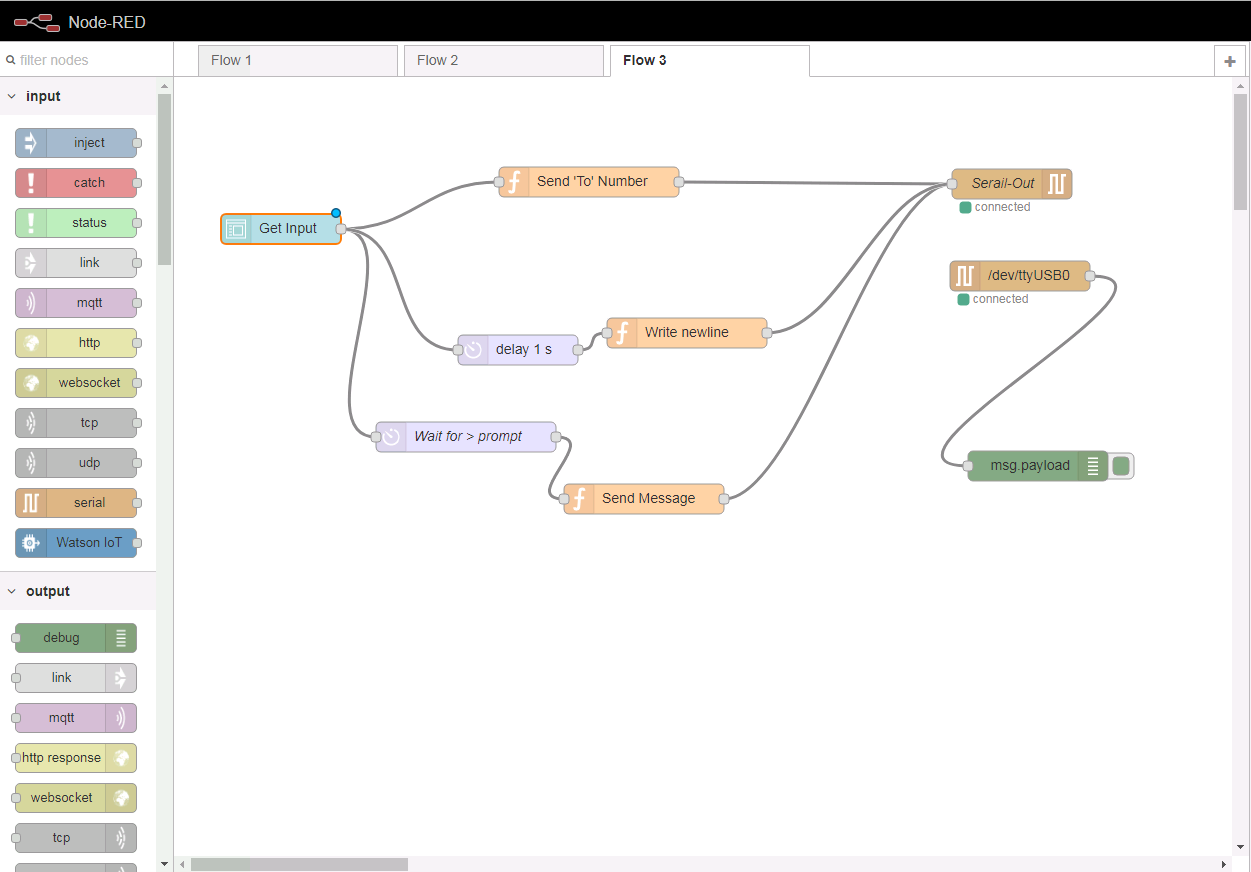}
\caption{A Node-Red interface that shows an illustrative Node-Red workflow using typical input and output components.}
\label{fig:node-red-interface}
\end{figure}

\subsection{Node-Red}
\label{subsec-node-red}

Node-Red\footnote{Node-Red: https://nodered.org/}, illustrated in Figure \ref{fig:node-red-interface}, is a low-code development tool designed for building event-driven applications by integrating hardware, APIs, and online services. It offers a browser-based flow editor that lets users configure application logic via a simple drag-and-drop interface. Built on Node.js, Node-RED’s lightweight, event-driven runtime is optimized for low-cost devices like the Raspberry Pi and scalable cloud environments. With its extensive library of nodes and compatibility with over 225,000 Node.js modules, Node-RED simplifies complex integrations and supports rapid deployment of IoT and web-based applications.

We adopted Node-Red in our proposed MoT approach for two reasons. First, Node-Red’s low-code interface enables rapid integration of diverse IoT devices and data sources. This ease of use streamlines the process of connecting sensors and cloud services within the MoT framework. We believe this can promote data flow management and reduce the need for extensive coding. Secondly, the event-driven model built on Node.js supports real-time data processing, which is essential for CoT applications. Its compatibility with a wide range of APIs and services also enhances the MoT approach's adaptability, enabling seamless scalability and responsiveness in handling IoT-driven events.

\section{Related Work}
\label{sec:trabalhos-relacionados}

Research on CoT applications has evolved along several interconnected strands that collectively shape the field's current state. These include: (i) \textit{formal modeling and verification approaches} that ensure behavioral correctness of IoT–Cloud systems, (ii) \textit{dataflow-centric and decentralized execution frameworks} designed to manage large-scale and distributed IoT data processing, (iii) \textit{model-driven engineering (MDE) solutions} that promote abstraction, automation, and systematic code generation, and (iv) \textit{deployment-oriented frameworks} that aim to reduce configuration burden and operational complexity in heterogeneous CoT environments. These lines of research are mature, yet each reflects distinct priorities and trade-offs.

Despite these advances, most existing approaches privilege either \textit{formal rigor over accessibility} or \textit{automation over structured model-driven design}. Formal solutions often provide strong guarantees but demand specialized expertise, limiting their applicability to broader developer communities. Conversely, automation-oriented and execution-focused frameworks streamline deployment but rarely treat modeling semantics as a first-class concern. Moreover, few works objectively integrate \textit{UML-based modeling}, low-code development principles, and executable automation in a unified, coherent manner. To analyze the literature rigorously and fairly, our review adopts explicit criteria, i.e., modeling paradigm, research focus, automation level, deployment support, and target users, enabling objective comparison rather than subjective assessment.

To this end, we present the procedures adopted to select our related studies (Section \ref{subsec-Study-Selection-Procedures}), analyze each selected study (Section \ref{subsec-analysis-related-work}), and then present a comparative analysis and research opportunities (Section 

\subsection{Study Selection Procedures}
\label{subsec-Study-Selection-Procedures}

The selection of related studies followed a structured process to ensure relevance, rigor, and alignment with this research's objectives. First, the authors' prior research experience in the Cloud-of-Things, Internet of Things, and Model-driven Engineering domains guided the identification of seminal and representative contributions. Second, a structured search strategy was applied using the string \textit{((``Cloud of Things'' OR ``Internet of Things'') AND UML AND ``Low code'' AND ``Model-based Development'')}. This query was executed in two widely recognized digital libraries---Google Scholar\footnote{Google Scholar: https://scholar.google.com/} and IEEE\footnote{IEEE Xplore: https://ieeexplore.ieee.org/Xplore/home.jsp}---to capture both academic impact and breadth of coverage. Third, only studies directly addressing themes aligned with our research focus were retained, particularly those that explored model-based development, automation, abstraction, and the feasibility of deployment in IoT or CoT systems. Works that merely cited these domains without engaging their core challenges were discarded. This three-step procedure ensured that the selected works address a research problem closely related to that investigated in this study and thus provide a meaningful basis for comparison.

This selection strategy is rigorous for three main reasons. First, it follows established methodological practices used and validated in prior peer-reviewed software engineering and IoT research, ensuring methodological soundness. Second, we carefully constructed the search string to reflect the core dimensions of our investigation (i.e., Cloud/IoT context, UML and model-based development, and low-code principles), reducing ambiguity and minimizing irrelevant results. Third, we focused on contributions published between 2020 and 2026, prioritizing work that reflects the latest advancements and technological constraints in the CoT ecosystem. We excluded literature reviews, surveys, and systematic mappings to focus exclusively on primary, solution-oriented contributions. Together, these procedures ensure the selected corpus is representative and tightly aligned with the research goals.

\subsection{Analysis of Related Work}
\label{subsec-analysis-related-work}

\textbf{Modeling paradigm.} Existing studies employ diverse modeling strategies to address the complexity of Cloud–IoT ecosystems. Hellal et al. \cite{rw01-hellal2024formal} adopt Event-B to formally specify and verify the structural, functional, operational, and behavioral properties of IoT–Cloud systems. In contrast, Fazel and Wainer \cite{rw03-fazel2023devs} employ the Discrete Event System Specification (DEVS) formalism to support model-driven simulation and analysis of IoT systems. Other studies emphasize dataflow modeling, as Zenoh-Flow \cite{rw02-baldoni2023dataflow}, organizing ML-powered IoT applications into decentralized data pipelines. Model-Driven Development (MDD) also underpins microservice-oriented modeling \cite{rw05-belguidoum2022mdmsd4iot}, ThingML-based development \cite{rw07-fortas2022application}, WoT-based frameworks \cite{rw06-de2020approach}, and cross-platform IoT domain-specific modeling languages (DSMLs) \cite{rw14-Corradini2023}. More recently, LocSys \cite{rw13-Panayiotou2025} has introduced a multi-DSL approach to support complex cyber-physical systems (CPS). Although these studies advance modeling abstraction and automation, they pursue different objectives, including formal verification, simulation, interoperability, and platform-independent development. Few explicitly integrate UML-based modeling, low-code execution, and automated deployment into a unified workflow accessible to developers with varying levels of expertise, which motivates the MoT approach.

\textbf{Research focus.} The literature addresses several distinct but complementary concerns. Works such as \cite{rw01-hellal2024formal,rw03-fazel2023devs} address correctness, verification, and behavioral assurance. Others prioritize system management, ML processing, and performance optimization in distributed IoT pipelines, as in \cite{rw02-baldoni2023dataflow}. Security-driven models are emphasized in \cite{rw04-sree2022mobile}, while architectural scalability and microservices structuring guide \cite{rw05-belguidoum2022mdmsd4iot}. \cite{rw09-hassan2021toward, rw10-terracher2024model} explicitly target deployment and distributed execution simplification, and \cite{rw06-de2020approach}. Meanwhile, advanced optimization and networking intelligence dominate \cite{rw10-Jiuting2025, rw11-Mohajer2024, rw12-Kuang2024}. LocSys \cite{rw13-Panayiotou2025} and X-IoT \cite{rw14-Corradini2023} enhance usability and portability by providing refined platform-independent abstractions. However, despite this breadth, most works do not explicitly seek to reduce development barriers for non-expert developers via low-code principles integrated with MDD, nor do they explicitly couple modeling semantics with executable, cloud-ready structures as MoT does.

\textbf{Automation level.} Automation appears with distinct emphases across studies. Full formal verification pipelines, as in \cite{rw01-hellal2024formal}, focus on correctness rather than execution automation. Automated ML data pipelines emerge in \cite{rw02-baldoni2023dataflow}, while simulation-driven automatic execution is supported in \cite{rw03-fazel2023devs}. Microservice generation is automated in \cite{rw05-belguidoum2022mdmsd4iot}, and ThingML \cite{rw07-fortas2022application} automates multi-platform code generation. Cross-platform deployment automation appears in \cite{rw09-hassan2021toward} and \cite{rw14-Corradini2023}. However, few tools automate end-to-end transformation from conceptual UML modeling to low-code executable artifacts, maintaining semantic continuity across steps. MoT explicitly addresses this blind spot by providing transparent transformation and gradual automation aligned with CoT deployment.

\textbf{Deployment support.} Deployment capabilities vary markedly. Simulation-oriented approaches \cite{rw03-fazel2023devs, rw07-fortas2022application} emphasize experimentation but require further effort for operationalization. Cloud deployment is explicit in \cite{rw01-hellal2024formal,rw06-de2020approach,rw09-hassan2021toward,rw10-terracher2024model}, while decentralized runtime support dominates in \cite{rw02-baldoni2023dataflow}. Performance-centric cloud execution is the primary concern of \cite{rw10-Jiuting2025, rw11-Mohajer2024, rw12-Kuang2024}. LocSys \cite{rw13-Panayiotou2025} provides execution pipelines but focuses on CPS rather than CoT execution. Although these studies advance cloud execution, most either assume expert users or introduce significant configuration complexity. MoT bridges this gap by explicitly connecting modeling constructs to deployable components using a structured, low-code execution backbone.

\textbf{Target users.} Most reviewed approaches implicitly assume experienced developers, formal method specialists, or IoT engineers. Event-B \cite{rw01-hellal2024formal} and DEVS simulation \cite{rw03-fazel2023devs} demand specialized expertise. ML-centric frameworks \cite{rw02-baldoni2023dataflow} and networking optimization works \cite{rw10-Jiuting2025, rw11-Mohajer2024, rw12-Kuang2024} are intended for advanced researchers and engineers. Microservice- and WoT-oriented solutions \cite{rw05-belguidoum2022mdmsd4iot, rw06-de2020approach, rw09-hassan2021toward} still require architectural understanding. Only a few explicitly prioritize accessibility and usability, notably LocSys \cite{rw13-Panayiotou2025}, which demonstrates strong user acceptance. Yet even these solutions rarely provide a streamlined, low-code path from UML modeling to concrete CoT deployment, which MoT aims to achieve.

In summary, the current literature shows maturity in modeling rigor, architectural design, and performance optimization, but it lacks comprehensive, low-code, model-driven solutions that explicitly reduce development barriers, preserve semantic continuity from models to execution, and support CoT deployment with accessibility as a first-class concern. These gaps define the core contribution space of our proposed approach.

\begin{sidewaystable}
\small
\centering
\begin{tabular}{|l|l|l|l|l|l|}
\hline
              & \textbf{Modeling} & \textbf{Research} & \textbf{Automation} & \textbf{Deployment} & \textbf{Target}\\ 
\textbf{Work} & \textbf{Paradigm} & \textbf{Focus} & \textbf{Level} & \textbf{Support} & \textbf{Users}\\ 
\hline

\textit{MoT approach} & \textit{UML modeling} & \textit{Automation + Streamline} & \textit{High} & \textit{Model-oriented + Cloud Agnostic} & \textit{Non-experts} \\ \hline

\cite{rw01-hellal2024formal} & Formal/Event-B & Verification & Medium & Cloud-oriented & Experts \\
\hline
\cite{rw02-baldoni2023dataflow} & Dataflow Framework & ML Pipelines & High & Decentralized IoT & Developers \\
\hline
\cite{rw03-fazel2023devs} & DEVS MDD & Simulation + Design & High & Simulation + Runtime & Experts \\
\hline
\cite{rw04-sree2022mobile} & Security Model & Secure Access & Low & Cloud & Practitioners \\
\hline
\cite{rw05-belguidoum2022mdmsd4iot} & SysML/MDD & Microservices & Medium & IoT Runtime & Developers \\
\hline
\cite{rw06-de2020approach} & MDE/WoT & Interoperability & Medium & Cloud/WoT & Developers \\
\hline
\cite{rw07-fortas2022application} & DSL (ThingML) & Code Gen + Testing & High & Multi-platform & Developers \\
\hline
\cite{rw09-hassan2021toward} & Declarative YAML & Distributed IoT Deploy & Medium & Cloud + Edge & DevOps Engineers \\
\hline
\cite{rw10-terracher2024model} & MDE & Sensor Data Exploitation & Medium & Cloud & Engineers \\
\hline
\cite{rw10-Jiuting2025} & Optimization Model & Network Performance & High & Satellite Cloud & Researchers \\
\hline
\cite{rw11-Mohajer2024} & DRL Framework & Edge Offloading & High & Edge Cloud & Researchers \\
\hline
\cite{rw12-Kuang2024} & MADL/GNN & Reliability + QoS & High & MANET Cloud & Researchers \\
\hline
\cite{rw13-Panayiotou2025} & Multi-DSL MDD & CPS Usability & Medium & Runtime Pipelines & Mixed Users \\
\hline
\cite{rw14-Corradini2023} & DSML MDD & Cross-Platform IoT & Medium & Platform-Independent Deploy & Developers \\
\hline
\end{tabular}
\caption{A comparative analysis of the selected studies.}
\label{tab:comparative-analise}
\end{sidewaystable}

\subsection{Comparative Analysis and Research Gaps}
\label{subsec:comparative-analysis}

Table \ref{tab:comparative-analise} presents our comparative analysis, drawing directly on the objective gaps identified rather than subjective perceptions. The synthesis of the selected works shows ongoing evolution in Cloud-of-Things engineering. However, it also reveals structural limitations that constrain development accessibility, architectural clarity, and execution feasibility. By analyzing studies against criteria such as modeling paradigm, research focus, automation level, deployment support, and target users, we identify three persistent research gaps that the current literature has not sufficiently addressed.

First, there is limited integration between UML-based modeling and low-code execution environments. Several solutions emphasize strong modeling capabilities \cite{rw01-hellal2024formal,rw03-fazel2023devs,rw05-belguidoum2022mdmsd4iot,rw14-Corradini2023}, while others privilege execution, deployment, and runtime efficiency \cite{rw02-baldoni2023dataflow,rw07-fortas2022application,rw09-hassan2021toward}. Yet, few approaches bridge these two perspectives to enable non-expert or mixed-expertise users to move seamlessly from conceptual models to executable CoT applications. Second, many works excel either in formal rigor and theoretical soundness \cite{rw01-hellal2024formal,rw03-fazel2023devs} or in automation and optimization \cite{rw10-Jiuting2025,rw11-Mohajer2024,rw12-Kuang2024}, but rarely balance abstraction with practical deployment capability, limiting their accessibility as engineering tools. Third, most solutions lack explicit end-to-end transformation pipelines that connect modeling constructs to deployable artifacts in a reproducible and extensible manner, particularly when heterogeneous cloud services and IoT platforms are involved.

The proposed MoT approach is designed to address these gaps. First, MoT introduces an engineering-focused solution centered on a UML profile tailored to CoT concerns, enabling structured design while maintaining familiarity for practitioners. Second, it defines a clear software architecture and implements a concrete tool-chain that automatically transforms models into platform-ready execution artifacts grounded in low-code principles. This moves beyond theoretical modeling by demonstrating operational feasibility. Third, MoT emphasizes practical validation. Beyond conceptual soundness, it is evaluated through a case study and user-centered assessment, providing qualitative evidence of usability, feasibility, and perceived usefulness, an aspect seldom addressed in comparable studies. 

\section{The Proposed Approach}
\label{sec:mot}

This section presents \textit{MoT (Model of Things}), a tool-supported, model-based approach that streamlines the development of Cloud-of-Things applications through low-code principles and UML model transformations. MoT emphasizes practicality and executability by tightly integrating modeling constructs with automated transformations, enabling a coherent path from design to deployment. We first provide an overview of the approach (Section~\ref{subsec_overview}), then introduce the proposed UML profile (Section~\ref{subsec_uml_profile}) and summarize MoT's key features (Section~\ref{subsec_key_features}). Next, we explain how model transformations are executed (Section~\ref{subsec_Transformations}) and describe the approach's underlying architecture (Section~\ref{subsec_architecture}). We also discuss key implementation aspects (Section~\ref{subsec_implementation_aspects}) and conclude by outlining the study's limitations (Section~\ref{subsec:limitations}).

\subsection{Overview of the Proposed Approach}
\label{subsec_overview}

Figure~\ref{fig:overview} presents an overview of our proposed 7-step process for streamlining CoT application development using low-code development and UML model transformations. The process comprises seven steps, each with specific inputs and outputs. Each step of this process is described as follows:

\begin{figure}[h!]
\centering
\includegraphics[scale=0.13]{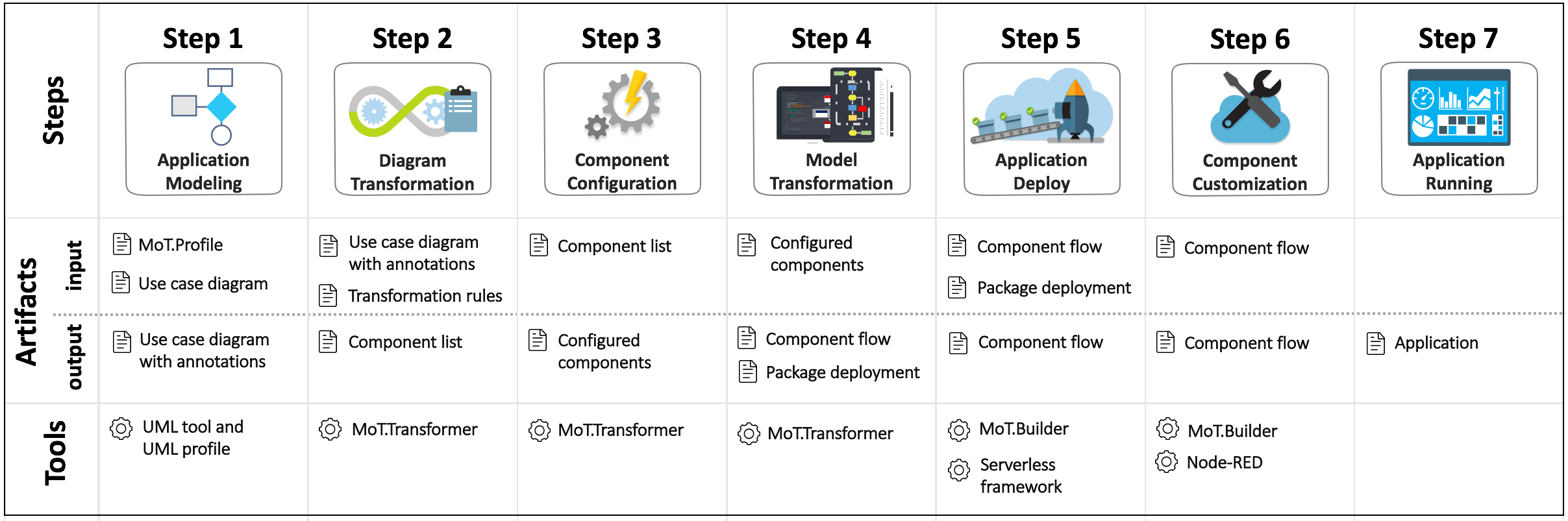}
\caption{An overview of the MoT process. It presents a seven-step, model-based software development process, from modeling and transformation to deployment and execution, using specific artifacts and tools to ensure a structured, efficient workflow.}
\label{fig:overview}
\end{figure}

\begin{figure}[h!]
\centering
\includegraphics[scale=0.15]{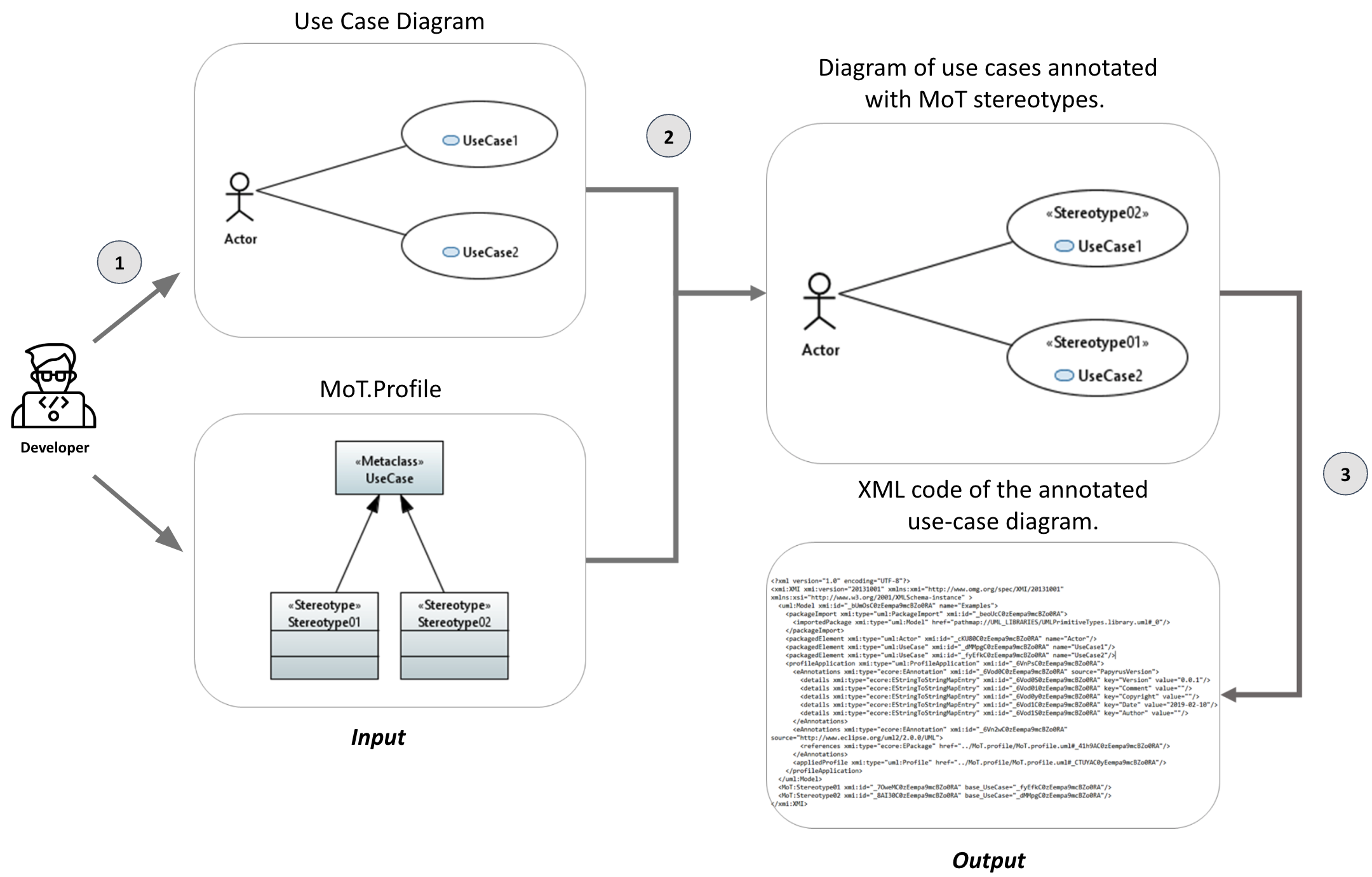}
\caption{An overview of the supported modeling process. Developers create UML use case diagrams (Step 1) and apply MoT stereotypes manually (Step 2). After the MoT approach generates an XML code of the annotated diagram (Step 3).}
\label{fig:step1-modeling}
\end{figure}

\textbf{Step 1: Application modeling.} This step aims at defining and specifying the requirements of CoT applications. Figure~\ref{fig:step1-modeling} shows the application modeling process. The application requirements are determined by creating UML use cases, a typical computation-independent model in model-based software development. The UML use cases can be created and refined as needed. It is an iterative process in which requirements are identified and refined through discussions among business specialists and technology experts. A use-case model shows how users (namely, actors) interact with the features of CoT applications, with application features, and with other actors. In this sense, the MoT process requires essential inputs, such as use cases, actors, and their relationships. The MoT users apply the stereotypes of the MoT profile (described in Section~\ref{subsec_uml_profile}) to customize the created use cases for the CoT domain. An annotated use case diagram is obtained from the use case diagram, with semantic information represented using stereotypes from the MoT profile. This model is then defined in XMI code, an OMG-standard XML-based information exchange format widely used by UML modeling tools. The XML code will be used as input for the next step.

\begin{figure}[h!]
\centering
\includegraphics[scale=0.18]{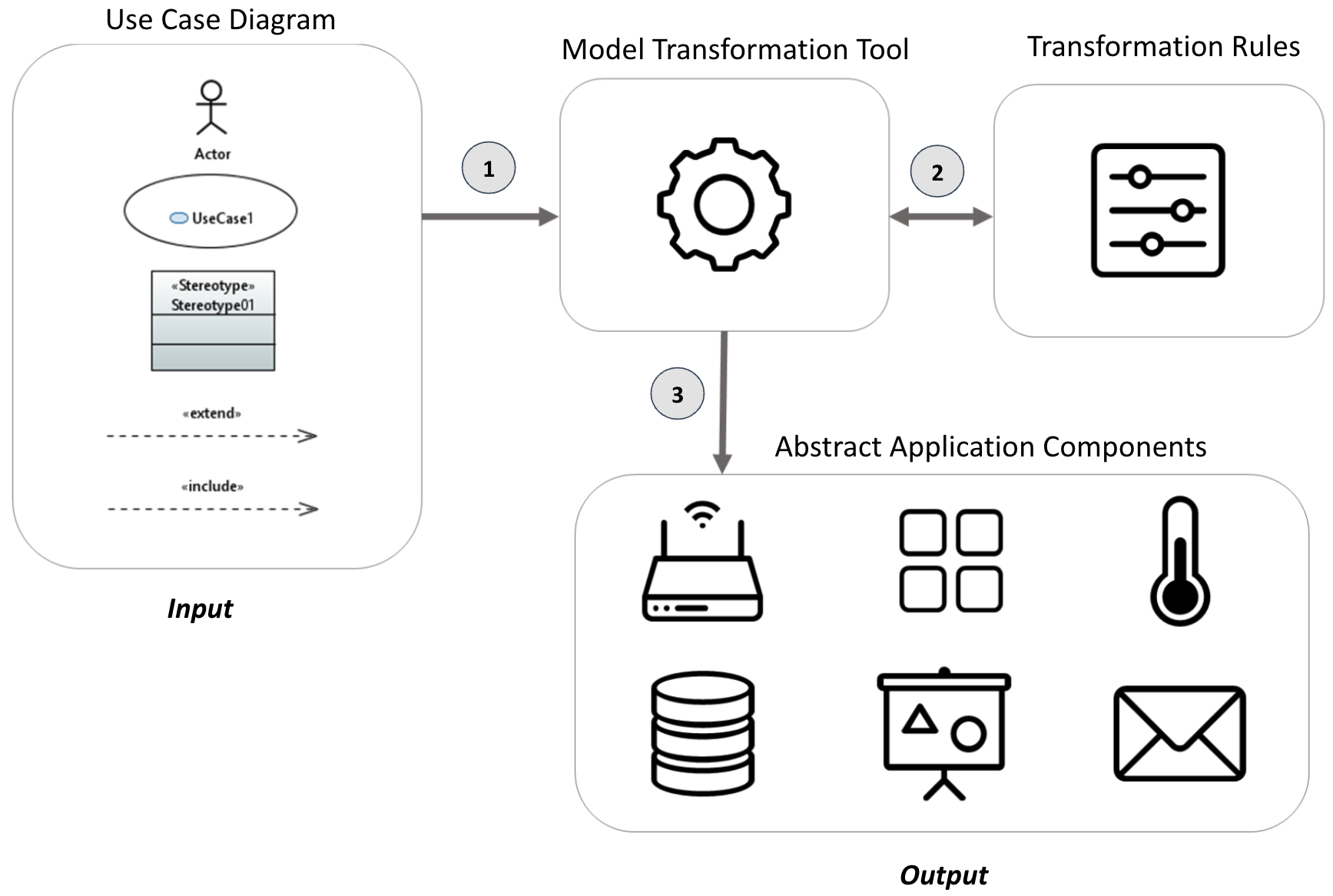}
\caption{An overview of the model transformation process supported. The MoT's model transformation tool receives a use case diagram as input (Step 1) and uses transformation rules (Step 2) to generate abstract application components as output artifacts (Step 3).}
\label{fig:step2-transformation}
\end{figure}

\textbf{Step 2: Diagram transformation.} With the application model developed in the previous step and represented in XMI format, this step initiates the first transformation of models in the MoT approach. The MoT.Transformer maps all the diagram elements and identifies stereotypes compatible with the UML profile. It then transforms these UML elements into abstract application components. Figure \ref{fig:step2-transformation} illustrates this transformation process. The elements in the use-case diagram, common diagram elements, and the \textit{MoT.Profile} elements used in the diagram are all mapped and analyzed. The key player in this process is the model transformation tool, which contains a set of transformation rules used to generate the application components. These application components represent specific elements and features of a CoT application. Some of these components may require additional configurations not covered in the use-case diagram.
 
 \textbf{Step 3: Component configuration.} It depends on user interaction to provide additional information for a particular application component. This information is specific to a particular application, such as database connection information, sensor signing/publishing information, user information, and social network passwords. These settings are necessary because the MoT user can only represent some of the required information through the UML diagram. In this configuration, if the user lacks data storage for application data from a particular service, such as a database service, the MoT approach can provision that service with cloud computing providers. This way, users can configure cloud computing services without technical expertise. The abstraction of technical details is one of MoT's characteristics. The simplification of the CoT application development process involves abstracting cloud computing services, configurations, and components, including software components, devices, and sensors in the context of IoT.

 \textbf{Step 4: Model transformation.} Automatic generation of platform-specific models through a generative approach to the MDE, which aims to transform models to models or models to code. Model transformation relies on information about the application's components, and the transformation tool translates component information and configurations into platform-specific components. These models represent the same application as those involved in previous activities, but they translate higher-level abstractions into platform-specific implementations. For example, this includes conversion types for platform-specific sizes and mapping the application elements to the platform programming model. In addition to transforming the components for a specific platform, this step generates a deployment package that includes the infrastructure configuration files required to deploy and execute the application in a cloud environment.

\begin{figure}[h!]
\centering
\includegraphics[scale=0.3]{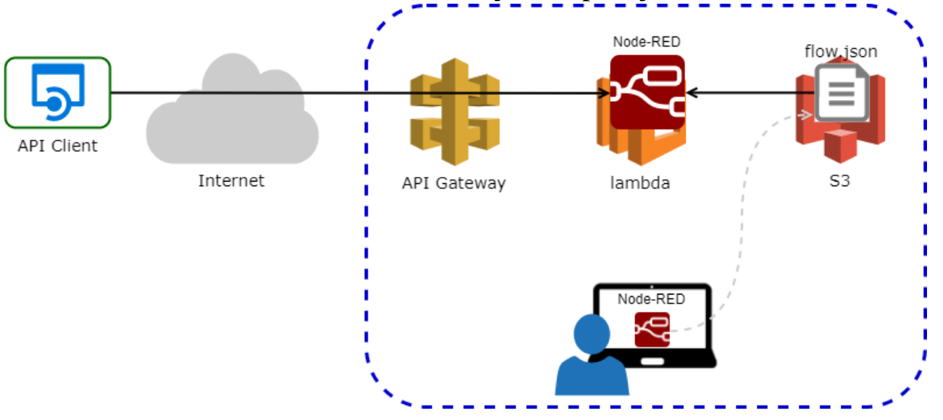}
\caption{An architecture in which an API client communicates via API Gateway to an AWS Lambda running Node-RED, which retrieves flow configurations from S3, enabling remote workflow execution and management.}

\label{fig:step5}
\end{figure}

 \textbf{Step 5: Application deployment.} After performing the required model transformations, the MoT approach prepares the application for deployment. In practice, this step establishes the cloud computing environment for running the CoT application. This cloud environment is based on the deployment package generated in the previous step. Figure~\ref{fig:step5} presents an infrastructure model created for the cloud computing provider Amazon Web Services (AWS): API Gateway acts as a ``gateway'' for applications to access data, business logic, or functionality in their back-end; Lambda allows code execution without provisioning or managing servers; and S3 is a service for storing files. Cloud computing providers, e.g., Amazon Web Services, Google Cloud Platform, and Microsoft Azure, are heterogeneous. Developers must fully understand this heterogeneity to push CoT applications to these providers. However, this understanding is hard to acquire, making it challenging to migrate CoT applications between providers. To mitigate this problem, this step facilitates CoT application delivery by abstracting the complexity introduced by heterogeneous cloud providers and automating the creation of a runtime environment for the CoT application. If the application requires customizations before delivery, this step will also prepare it for these customizations. This delivery customization is done in the next step.
 
 \textbf{Step 6: Delivery customization.} Customizations are sometimes required before delivering a CoT application, or mandatory data must be collected to enable the delivery of CoT applications. This step adapts the application components to a specific platform. For example, sensitive user information, such as login credentials, is collected to deploy an application component to a cloud computing provider. This step implements platform-specific modifications and obtains sensitive user information that is not considered in Step 4. The next step starts if no customization or sensitive information is needed.

\textbf{Step 7: Application delivery.} This step ends the development cycle of a CoT application. It focuses on delivering the CoT application itself. In this sense, this step uses delivery procedures to ensure the features of CoT applications are accessible to their users. The MoT user can run the delivered CoT application locally for testing and approval. Afterward, it can be migrated to a cloud computing environment, thereby ending the development process.

\subsection{UML Profile}
\label{subsec_uml_profile}

UML~\cite{UML2017,kleinner2017modeling} is a general-purpose language that can model a wide range of application domains; in certain situations, it cannot model applications defined in specific domains (\textit{e.g.}, financial applications, time applications, real multimedia applications, among others). Thus, the language needs extension. The UML can be extended in two ways. The first form is through a lightweight extension. A UML profile is created to add semantic annotations to the UML metamodel, adapting it for specific platforms (such as Spring Boot, J2EE, and EJB) or application domains (such as finance and telecommunications) by adding stereotypes, tag values, and constraints. The second one represents a heavyweight extension. The UML metamodel is modified non-conservatively, altering the standard semantics of its elements.

The MoT approach proposes a UML profile, known as \textit{MoT.Profile}, which comprises a set of stereotypes to support the representation of CoT application concepts \cite{oliveira2008}. A stereotype is an extension of an existing UML metaclass or another stereotype, possibly defining additional attributes. Table \ref{tab:mot-profile} presents the \textit{MoT.Profile}. In total, the MoT.Profile defines nine stereotypes organized into five categories. Each category groups stereotypes by their intended use. For example, \textit{SensorSubscribe} and \textit{SensorPublish} fall under the IoT category because they connect to an IoT sensor.

\begin{table}[ht]
\footnotesize
\centering
\begin{tabular}{|l|l|l|}
\hline
\footnotesize
\textbf{Category} & \textbf{Stereotype} & \textbf{Description} \\ 
\hline
IoT  & $\ll$SensorSubscribe$\gg$ & Monitors data from an IoT sensor by connecting to \\ & & an MQTT Broker and signing messages from  \\ 
& & a specific topic. \\ 
\hline
IoT  & $\ll$SensorPublish$\gg$ & Sends data to an IoT sensor by connecting to  \\ & & an MQTT Broker and posting messages to \\ 
& & a specific topic. \\ 
\hline

Storage & $\ll$DatabaseSave$\gg$ & Establishes a connection to a Database server \\ 
& & to save data. \\ 
\hline

Dashboard & $\ll$DashboardGauge$\gg$ & Displays information in the form of a gauge on \\ & & a Dashboard page. \\ 
\hline
Dashboard & $\ll$DashboardChart$\gg$ & Displays information in the form of a chart on \\ & & a Dashboard page. \\ 
\hline
Emotiv BCI & $\ll$FacialExpression$\gg$ & With a set of Emotiv BCI tools to get brain-computer \\ 
& & interface integration to connect a facial expression. \\ 
\hline
Emotiv BCI & $\ll$MentalCommand$\gg$ & Using the Emotiv BCI (Brain-Computer Interface)  \\ 
& & toolkit allows to create brain-computer interface  \\ 
& & integration to capture the value of a mental command \\ 
& & from a profile. \\ 
\hline
Social & $\ll$TwitterPost$\gg$ & Connects to a Twitter account to post a post \\ 
& & on the social network. \\ 
\hline
Social & $\ll$SendEmail$\gg$ & Connects to an SMTP server to send e-mail messages. \\ 
\hline
\end{tabular}
\caption{A list of MoT.Profile stereotypes.}
\label{tab:mot-profile}
\end{table}

\subsection{Key features of MoT}
\label{subsec_key_features}

The key features of MoT that are provided are listed as follows:

\begin{enumerate}

\item{\textbf{Abstraction of IoT components.}} The proposed approach abstracts the complexity of the IoT components in UML models, which allows even those who do not have programming knowledge to be able, for example, to develop an application to connect to sensors and/or services offered by social networks, cloud computing, and APIs from other service providers.

\item{\textbf{Automated source code generation.}} Through a tool for transforming models into text, the approach generates the source code of the application based on the elaborated models. This feature enhances team productivity by enabling the reuse of models to build multiple applications and thereby ensuring higher-quality software.

\item{\textbf{Setting up the environment.}} Based on the components used to build the application, the approach realizes the creation and configuration of the environments in cloud computing providers to support the development of the CoT application. As such, the approach abstracts deployment details, reducing the complexity of building native cloud applications that fully leverage cloud computing benefits such as scalability and pay-as-you-go.

\item{\textbf{Extensibility.}} By creating new stereotypes and corresponding transformation rules, it is possible to add new features to the approach and thereby extend the use of the approach to other domains.

\end{enumerate}

The integration between UML profiles and Node-Red in MoT is not conceived as a simple engineering convenience, but as a principled architectural bridge that aligns platform-independent modeling abstractions with executable, low-code artifacts. The MoT UML profile defines a stable semantic layer for CoT concepts, enabling developers to consistently and reusably express system intent, interaction structures, and cloud behaviors. Node-RED then acts as an execution substrate in which these semantics are materialized as deployable flows. This structured mapping preserves architectural intent, reduces configuration effort, and mitigates the knowledge barrier for developers, while maintaining traceability from high-level models to executable implementations.

Although our implementation uses Node-Red as the primary execution platform, MoT was designed to remain general and reusable across heterogeneous IoT ecosystems. The stereotype-to-component mappings are declarative and repository-based, enabling extension or replacement without altering the conceptual model. This separation of concerns fosters portability and supports architectural evolution, enabling MoT to adapt to alternative cloud infrastructures and IoT development environments. By formalizing the bridge between model semantics and low-code execution, MoT moves beyond ad hoc platform use and provides a systematic, model-driven foundation for CoT application development.

\subsection{Transformations}
\label{subsec_Transformations}

\begin{figure}[h!]
\centering
\includegraphics[scale=0.17]{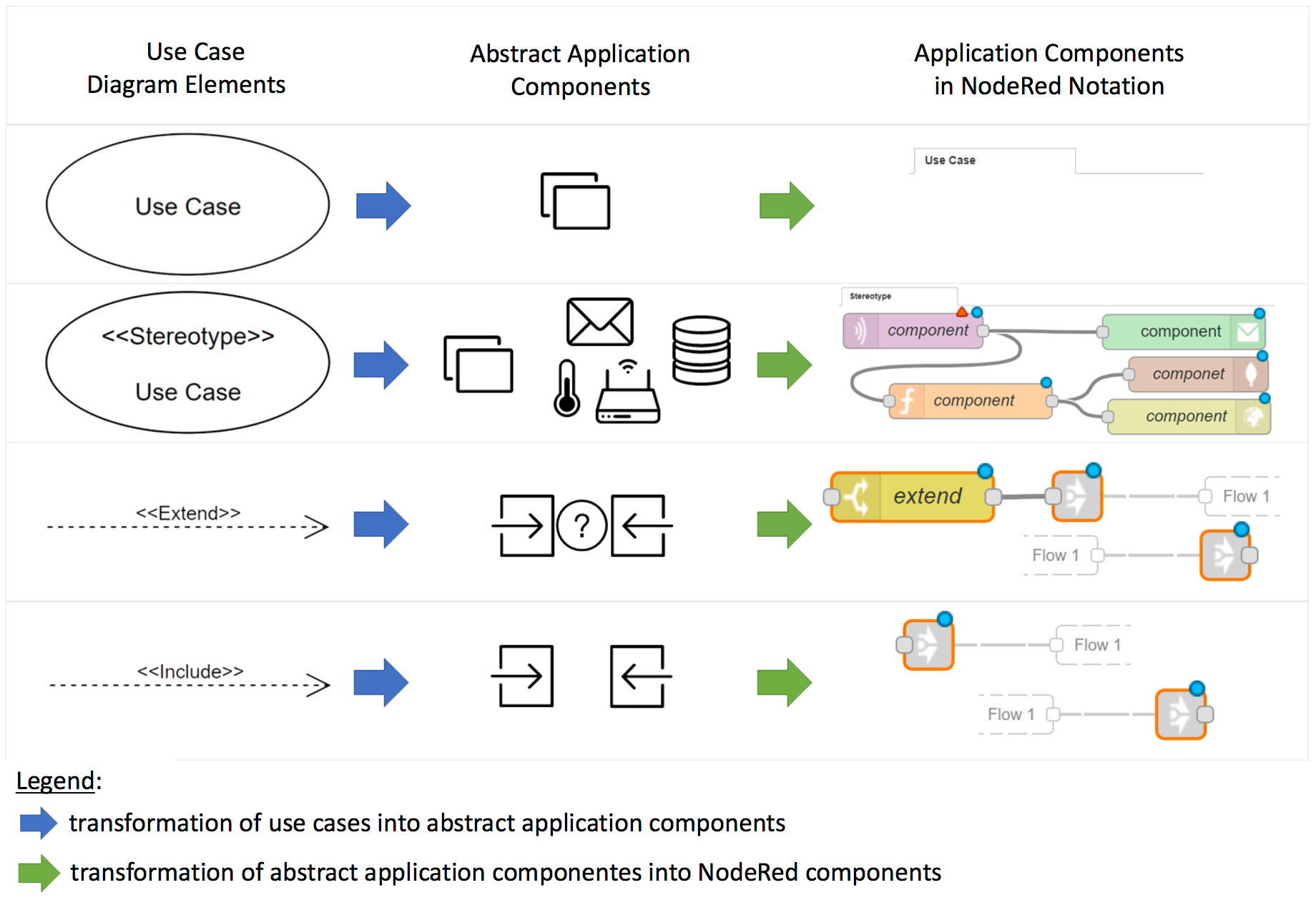}
\caption{Illustration of the mappings performed to support the transformation of use-case diagrams into Node-RED components.}
\label{fig_transformation}
\end{figure}

Figure~\ref{fig_transformation} illustrates the mappings used to transform use-case diagrams into Node-RED components. During the MoT approach process, there are two moments where model transformations occur (Step 2 and Step 4):

\begin{enumerate}

\item{Diagram transformation.} It occurs from analyzing the elements in the use case diagram and the stereotypes used in the UML \textit{MoT.Profile}. For each element and/or stereotype identified in this transformation, MoT performs a specific mapping of that element to an application component.

\item{Models transformation.} This occurs after the configuration steps of the application components. In this transformation, these components are transformed into components of the application's target platform.

\end{enumerate}

The MoT approach does not aim to establish a formal mathematical or algebraic foundation for model-driven development. Instead, it adopts a pragmatic, engineering-oriented strategy whose primary goal is to reliably transform models into executable systems through structure and automation. The tool expresses transformation behavior, execution constraints, and architectural guarantees concretely through well-defined transformation rules and controlled processing steps, rather than formal grammars or symbolic semantics. In particular, MoT operationalizes model semantics using structured JSON descriptors and a transformation pipeline that interprets UML stereotypes and transforms them into executable Node-RED flows. This design emphasizes repeatability, ensuring that identical models consistently produce functionally equivalent deployments, while also supporting extensibility through rule-based integration of new stereotypes and components.

\begin{figure}[h!]
\centering
\includegraphics[scale=0.13]{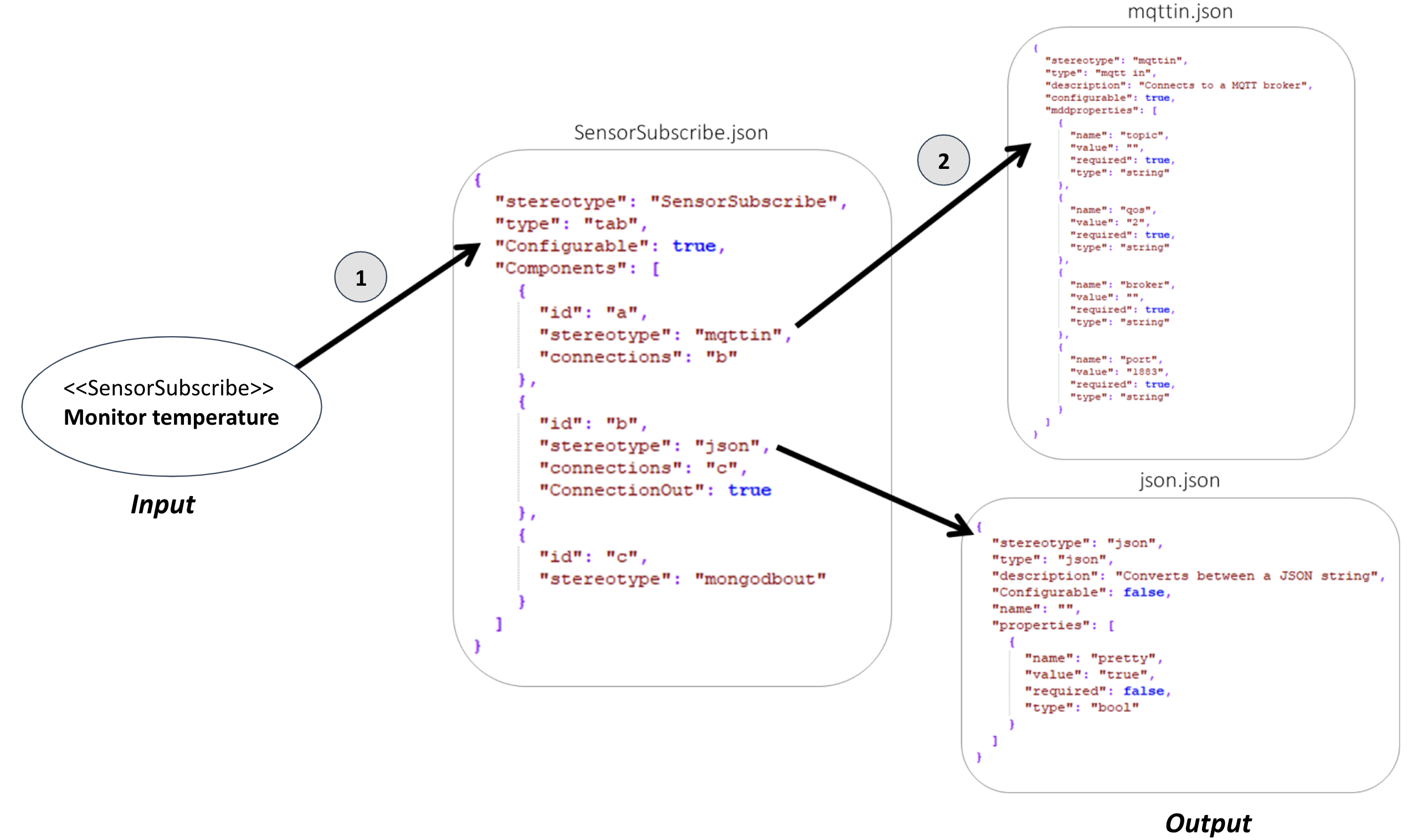}
\caption{Transforming the use case Monitor temperature containing $\ll$SensorSubscribe$\gg$ stereotype into components. The MoT receives a use case, labeled with a stereotype, as input, and accesses the $\ll$SensorSubscribe$\gg$ stereotype mapping (Step 1). Next, it applies transformation rules to generate components as output artifacts (Step 2).}
\label{fig_transformation_rule}
\end{figure}

Transformation rules guide the generation of executable artifacts from the modeled diagrams. During this process, for each stereotype identified, MoT retrieves the corresponding JSON specification from a repository of transformation descriptors. The MoT.Transformer then interprets this file to determine whether the stereotype maps to a single component or to a composite structure formed by multiple components. For composite stereotypes, the rule evaluation proceeds recursively until the process reaches a final component that contains only executable properties. These properties capture the essential configuration parameters required later in the Component Configuration step. This mechanism not only drives the execution pipeline but also reinforces extensibility: to introduce new functionality, developers only need to define new JSON descriptors and associate them with new stereotypes in the MoT profile. Figure~\ref{fig_transformation_rule} illustrates this process for the $\ll$SensorSubscribe$\gg$ stereotype.

Although MoT does not pursue numerical optimization or formal performance modeling, it operates in an environment where performance considerations are relevant. Therefore, MoT preserves architectural traceability between model constructs and execution behavior, enabling developers to reason about key characteristics of CoT applications, such as latency, payload behavior, deployment delay, and event frequency. Rather than constraining performance by design, MoT provides a structured execution basis for performance tuning, evaluation, and future performance-aware extensions. In doing so, MoT shows that model-driven abstractions can be transformed into working CoT applications with minimal manual intervention, while still enabling future research to explore optimization, formal semantics, and advanced performance strategies on this practical foundation.

\subsection{Architecture}
\label{subsec_architecture}

\begin{figure}[h!]
\centering
\includegraphics[width=0.9\textwidth]{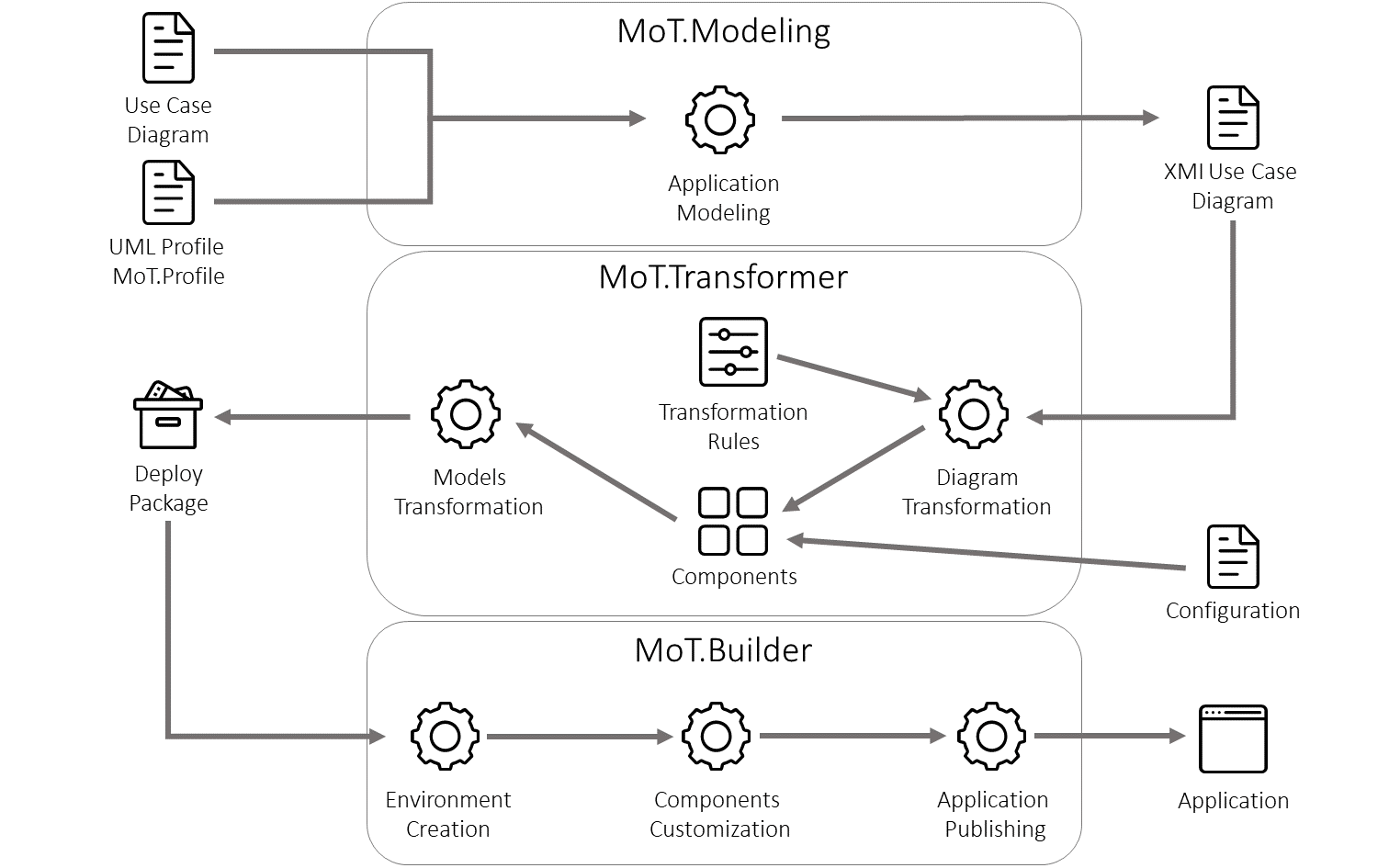}
\caption{Overview of the MoT architecture, illustrating the three core modules --- MoT.Modeling, MoT.Transformer, and MoT.Builder --- along with their interactions in modeling, transforming, and deploying CoT applications.}
\label{fig:arquitetura}
\end{figure}

Figure \ref{fig:arquitetura} shows the MoT approach's architecture and its three independent modules: \textit{MoT.Modeling}, \textit{MoT.Transformer} and \textit{MoT.Builder}. Each module follows well-known software design principles \cite{richards2020,martin2018clean}, including the single-responsibility and open-closed principles.

\begin{enumerate}
\item \textbf{MoT.Modeling:} This module models the CoT application. It takes as input a Use Case diagram and MoT Profile specifications. To support modeling, it integrates widely used industry tools, such as Enterprise Architect and Papyrus. The tool extends UML using UML profiles, an established extension mechanism. For example, an MoT Profile can define specific IoT components, such as sensors and actuators, within a standard UML diagram.  

\item \textbf{MoT.Transformer:} This module analyzes, configures, and transforms models. First, it applies transformation rules to extract application components. Next, it configures these components with the required information, for example, by assigning a cloud storage service to data-collecting sensors. Finally, it generates a deployment package, which the next module processes. For instance, a UML model of a temperature sensor is transformed into a deployable cloud function with its corresponding API endpoint.  

\item \textbf{MoT.Builder:} This module creates execution environments, customizes components, and publishes the application. It automates deployment using tools such as Docker and Kubernetes, starting from the generated deployment package. For example, after transforming the model, MoT.Builder can create a containerized application and deploy it to a cloud service such as AWS or Azure.

\end{enumerate}

To clarify MoT's architecture, we explicitly describe how the approach structures its transformation, execution, and deployment pipeline. MoT separates concerns into three well-defined layers (Figure \ref{fig:arquitetura}). The \textbf{modeling layer} captures IoT and cloud concepts using UML stereotypes, preserving semantic intent in a technology-agnostic manner. The \textbf{transformation layer} codifies these semantics into executable artifacts through organized, rule-based mappings, associating each stereotype with reusable components and well-defined responsibilities. This modular design clarifies execution behavior, improves maintainability, and supports consistent deployment across devices. The current literature recommends modularizing modern applications \cite{urdangarin2021}. Finally, the \textbf{deployment layer} automates generating and configuring target artifacts, ensuring consistency during execution and reducing human intervention. Although the current implementation targets a specific runtime platform, the layered organization, rule-based mappings, and stereotype-driven semantics provide a clear path toward extensibility across heterogeneous cloud environments.

\subsection{Implementation aspects}
\label{subsec_implementation_aspects}

This section outlines the technologies that were used to implement the proposed approach. We seek to explain how MoT Builder\footnote{MoT Builder: https://github.com/cristianowelter/MoT.Builder}, 
MoT Transformer\footnote{MoT Transformer: https://github.com/cristianowelter/MoT.Transformer}, and MoT Profile\footnote{MoT Profile: https://github.com/cristianowelter/MoT.Profile} were implemented. We implemented the MoT approach in C\# using ASP.NET and Visual Studio. The following section discusses the technologies used to implement each MoT module.

\textbf{MoT.Profile.} The UML profile of the approach was crafted using Eclipse Papyrus\footnote{Papyrus: https://www.eclipse.org/papyrus}, an Eclipse-based free software UML modeling environment that supports creating UML profiles.

\textbf{MoT.Modeling.} This module aims to generate the models of the application to be created. This modeling can be performed with tools widely used in the software industry, such as Eclipse Papyrus, Astah, and Modelio, among others available on the market. The tool supports the UML language and extends it via UML profiles, which extend the traditional UML standard. It also exports templates in accordance with the XMI standard, an OMG standard for XML-based information exchange. The main responsibilities of this tool are:

\begin{enumerate}
\item{Modeling the application.} Through the use of UML and the UML profile of the proposed approach, the requirements are formalized in models that will be represented in a diagram.
\item{Exportation of the models.} From the created models, the information in the diagrams is exported to XMI files, which the transformation module uses.
\end{enumerate}

\textbf{MoT.Transformer.} It provides functionality for managing applications created by the approach. For this, a tool was implemented that has the responsibility of interpreting the XMI files generated by the UML models in the modeling step, as well as performing the validation and configuration of the components, and at the end of the configuration process, carrying out the transformation of this information to a file that will be used by the next tool. This tool consists of a Web application that separates content (HTML), presentation (CSS 3), and functions (JavaScript). The most popular front-end component library, Bootstrap, has been used to create responsive Web projects. Bootstrap provides user-friendly, responsive interfaces to web applications. The approach used plugins created with jQuery\footnote{https://jquery.com/}, an open-source JavaScript library designed to simplify client-side scripting. The tool was implemented using Visual Studio Community, the C\# programming language, and ASP.NET Core, a free, cross-platform, high-performance software framework for creating modern cloud applications. To store the information, the tool uses a Microsoft SQL Server database. The main responsibilities of this tool are:

\begin{enumerate}
\item \textbf{Interpretation of the UML model.} The tool can interpret the XMI file exported by the UML modeling tool and, using the rules defined by the approach's UML profile, generate the pre-defined application components to be created.
\item \textbf{Components configuration.} Based on the components identified in the interpretation of the UML model, the tool asks the user for the necessary, specific information for each component. This information will complement the UML model. This configuration distinguishes one application from another because the same UML model can be used to create multiple applications, but these settings make the application unique.
\item \textbf{Models transformation.} After the configuration of all the components of the application to be generated, the tool performs the export of the information collected to a deployment package, which contains the information needed to build the application through the \textit{MoT.Builder} module.
\end{enumerate} 

\begin{figure}[h!]
\centering
\includegraphics[scale=0.18]{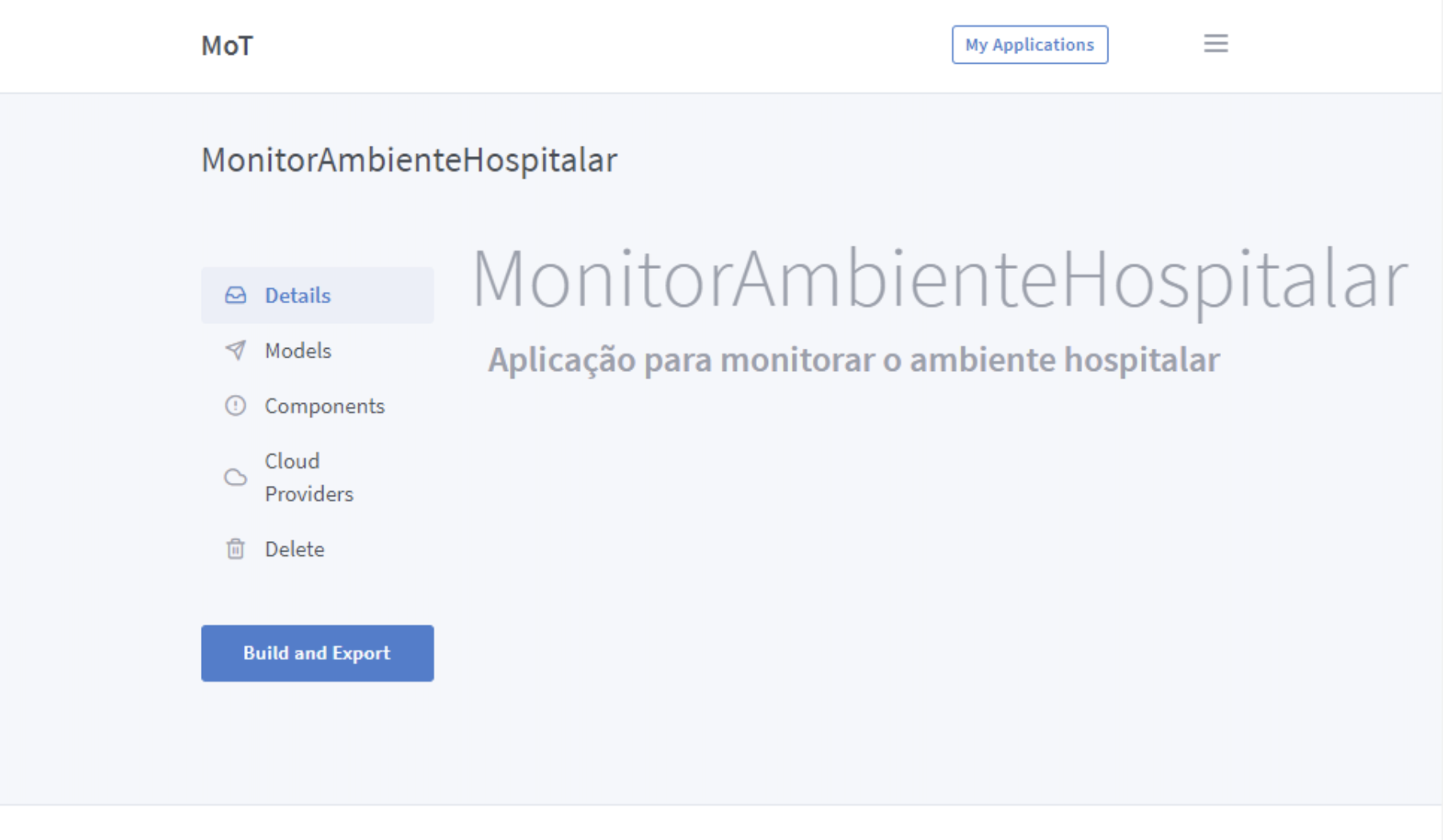}
\caption{Interface for application management of the MoT.Transformer tool.}
\label{fig_mot_transformer}
\end{figure}

Figure~\ref{fig_mot_transformer} shows the application management interface of the implemented tool. It is possible to identify the options for managing a CoT application; these options are: \textit{Details}: displays the details of the application; \textit{Models}: option to import the models created from the modeling step; \textit{Components}: option to access the application's component list and configure them; and \textit{Cloud Providers}: option to configure the data of a cloud computing service provider to create the infrastructure for execution.

\textbf{MoT.Builder.} The last module comprises executing the application, customizing the components, and ending its publication. This step requires two tools, both running on the Node.js platform, enabling JavaScript applications to run on the server. The first tool is Node-Red, a programming tool for connecting hardware devices, APIs, and online services, with a browser-based editor for connecting components and generating a stream. The Node-RED interface displays the application flow generated in the previous step, allowing the user to run the application, identify implementation issues, and, if necessary, customize the generated components. The second tool to be used in this step, and which also runs on the same platform, is the Serverless Framework\footnote{https://www.serverless.com/}, a toolkit for deploying serverless architectures across any cloud hosting provider that automates infrastructure configuration. The main responsibilities of these tools are:

\begin{enumerate}

\item{Execution of the application.} Through the Node-RED platform, you can run the generated application even outside a production environment; this helps identify flow issues, configuration problems, and application architecture failures.

\item{Customization of components.} If required, you can edit the application components and their settings via the Node-RED interface. We can also extend the application's functionality by adding new components, functions, and services. Still, these changes require some technical knowledge of JavaScript and familiarity with the platform.

\item{Application deployment.} After verifying and validating the application's functionalities, it can be deployed through the Serverless Framework tools. This process is automated and provides a link to access the generated application.

\end{enumerate}

\subsection{Limitations}
\label{subsec:limitations}

The current version of MoT adopts a deliberately engineering-oriented design, prioritizing feasibility and demonstrable automation over full formal expressiveness and adaptive transformation intelligence. Transformation logic is implemented via predefined mappings, which validate the approach's technical viability but constrain flexibility. Currently, MoT lacks dynamic adaptation or configurable transformation strategies that automatically adjust to alternative architectural choices, unforeseen modeling patterns, or platform-specific variations. Likewise, although we clarify the transformation process's internal behavior, the approach does not yet formalize its semantics or provide mathematically grounded validation mechanisms. These constraints suggest that MoT is a feasible, low-code proof of concept rather than a fully configurable, formally verified transformation framework.

Furthermore, the current implementation offers limited fault tolerance and operational resilience. MoT does not incorporate automated rollback procedures, robustness mechanisms to handle partial failures, or recovery workflows for incomplete model configurations or incorrect deployments. Error handling relies primarily on user intervention rather than autonomous platform support. In addition, this evaluation does not include quantitative performance indicators such as transformation latency, scalability under load, or execution failure rates. These omissions reflect the study’s focus on feasibility and usability rather than exhaustive performance benchmarking. We explicitly acknowledge these restrictions to make the operational boundaries of MoT transparent and to guide readers regarding its current maturity and intended use.

\textbf{Problem formulation and evaluation scope limitations.} A further limitation of this study concerns the scope and nature of its problem formulation and evaluation. The present work adopts a descriptive, engineering-oriented perspective, prioritizing feasibility and demonstrability over a formal optimization framing. Accordingly, we do not explicitly define measurable optimization objectives or constraints, such as reducing deployment latency, increasing automation coverage, reducing cloud costs, or improving bandwidth efficiency. Likewise, although the mapping between UML requirements and executable artifacts is systematic and rule-based, it has not yet been formally evaluated for correctness, coverage, or completeness. The study also does not quantify reductions in developer effort or provide baseline comparisons against alternative approaches. These omissions limit the ability to substantiate claims regarding efficiency gains and automation benefits with empirical metrics. We recognize these constraints and explicitly acknowledge that a more formal problem definition, benchmark-driven validation, and quantitative assessment remain essential directions for future work.

\textbf{Evaluation scope and external validity limitations.} Although the proposed approach demonstrates feasibility, we acknowledge essential limitations regarding the breadth and depth of its evaluation. The current study does not measure scalability, performance behavior under different workloads, or resilience in multi-configuration stress scenarios. Its empirical assessment relies on a single application domain and a constrained experimental setting, limiting external validity and reducing our ability to generalize the findings. We also recognize challenges in recruiting participants with expertise in both modeling and CoT technologies, the dependence on Node-Red as a demonstration platform, and the absence of strong comparable MDE tools for fair benchmarking. For these reasons, this work should be understood primarily as a feasibility and proof-of-concept investigation rather than a comprehensive validation study. Nonetheless, it establishes a necessary foundation by showing that the approach is technically executable end-to-end and meaningful to knowledgeable practitioners. Future work will address these limitations through multi-scenario case studies, quantitative benchmarks, comparative experimentation with alternative MDE solutions, and evaluation across heterogeneous IoT/cloud environments.

\section{Evaluation}
\label{sec_evalution}

This section presents the key decisions underlying the experimental procedures used to evaluate the proposed approach. We evaluated our approach through a case study developing a real-world CoT application (Section~\ref{subsec_case_study}). In addition, we assess the MoT approach with respect to users' perceptions of usability and acceptability (Section~\ref{subsec_TAM}). Our evaluation methodology adhered to practices established in previously published articles and followed widely known guidelines \cite{wohlin2012experimentation,d2020sw,farias2015evaluating,rubert2022effects}.

\subsection{Case study for Monitoring of Hospital Environments}
\label{subsec_case_study}

\textit{Scenario.} Continuous monitoring is essential in hospital environments to collect biometrics \cite{menzen2021}, as even minor failures can jeopardize lives. Regulations specify the environmental conditions required for patient care, vaccine storage, and medication preservation, which require precise temperature and humidity control. However, many hospital units still rely on manual monitoring. Automating this monitoring process presents a robust solution to these challenges. Key features include automated analysis, graph-based visualization, and customizable alert mechanisms.
The system's operational framework involves collecting, storing, and processing IoT sensor data for continuous monitoring and informed responses. From this scenario, we derive the following functional requirements for the proposed application:

\begin{itemize}
    \item Monitor environmental temperature.
    \item Store temperature data.
    \item Generate graphical representations of temperature data.
    \item Issue alerts for critical temperature variations.
\end{itemize}

We developed a CoT application using the proposed approach to fulfill the requirements of the hospital monitoring scenario. We detail the step-by-step process as follows:

\subsubsection{Step 1: Application modeling}

\begin{figure}[!h]
\centering
\includegraphics[width=0.8\textwidth]{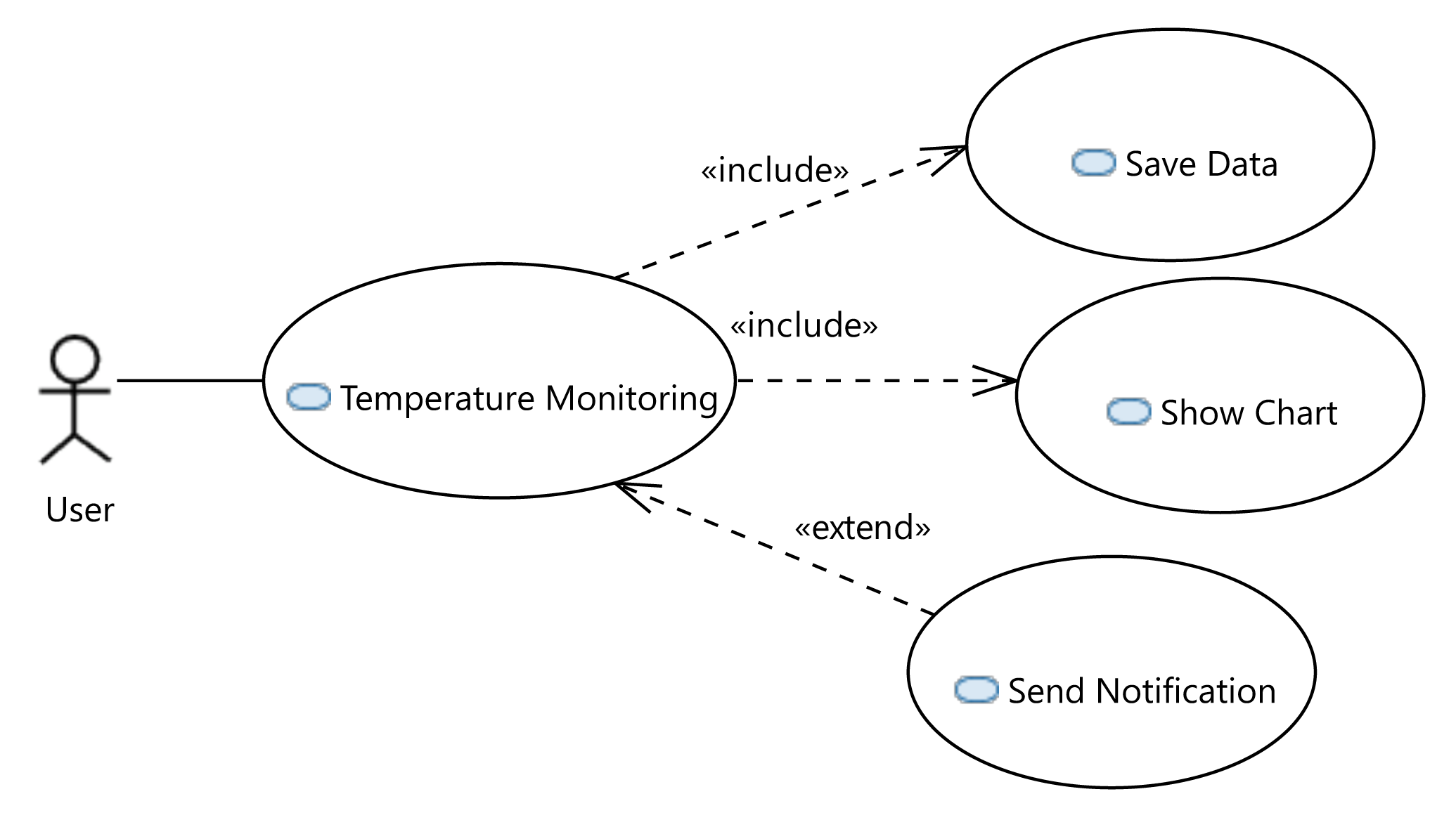}
\caption{Illustration of a use case diagram for temperature monitoring. The user monitors temperature, including data storage and graph visualization. Notifications extend monitoring, triggering alerts when predefined conditions are met.}
\label{fig:ec1-caso-uso}
\end{figure}

Figure \ref{fig:ec1-caso-uso} illustrates a UML use case diagram for a temperature monitoring system with core functionalities and their relationships. We used the open-source tool Eclipse Papyrus to elaborate on it. The primary actor, labeled as \textit{User}, interacts with the central use case, \textit{Temperature Monitoring}, which encompasses three related functionalities:

\begin{itemize}
    \item \textbf{Save Data}: The \textit{Temperature Monitoring} use case includes this functionality, represented by an \texttt{<<include>>} relationship. This indicates that data saving is a mandatory sub-process whenever temperature monitoring occurs. It ensures that all monitored temperature data is recorded persistently.

    \item \textbf{Show Chart}: Another \texttt{<<include>>} relationship links the \textit{Show Chart} functionality to the \textit{Temperature Monitoring} use case, indicating that generating and displaying charts is an essential part of the monitoring process. This function provides users with a visual representation of temperature data over time, enhancing the system's usability.

    \item \textbf{Send Notification}: The \textit{Send Notification} functionality is associated with \textit{Temperature Monitoring} through an \texttt{<<extend>>} relationship. This relationship implies that notifications are sent only under specific conditions, such as exceeding temperature thresholds. It allows the system to provide timely alerts without requiring every monitoring instance to trigger a notification.
\end{itemize}

\begin{figure}[!h]
\centering
\includegraphics[width=0.8\textwidth]{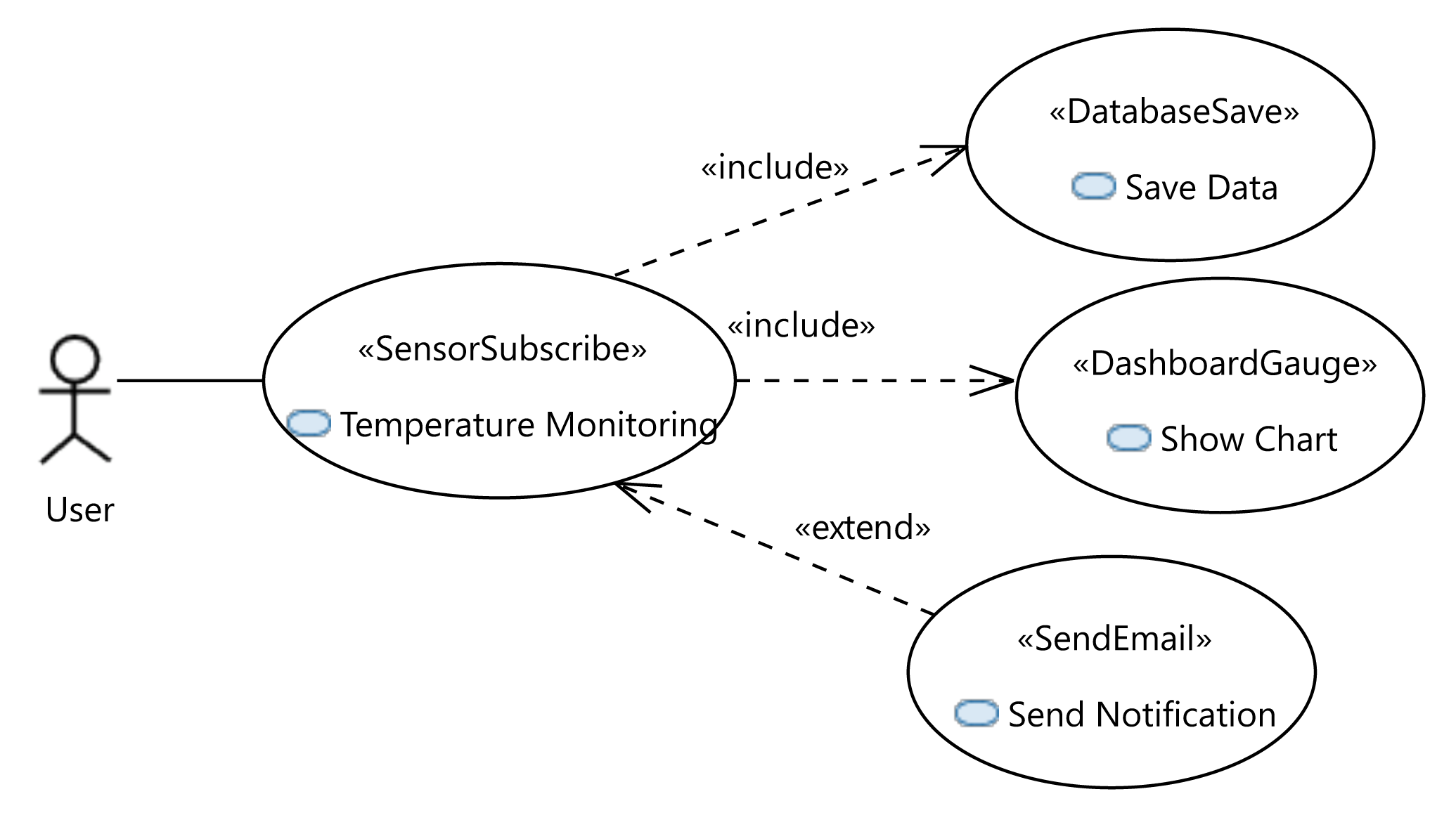}
\caption{Detailing the temperature monitoring use case using the proposed MoT.Profile stereotypes.}

\label{fig:ec1-caso-uso-estereotipos}
\end{figure}

\begin{figure}[h!]
\centering
\includegraphics[width=1\textwidth]{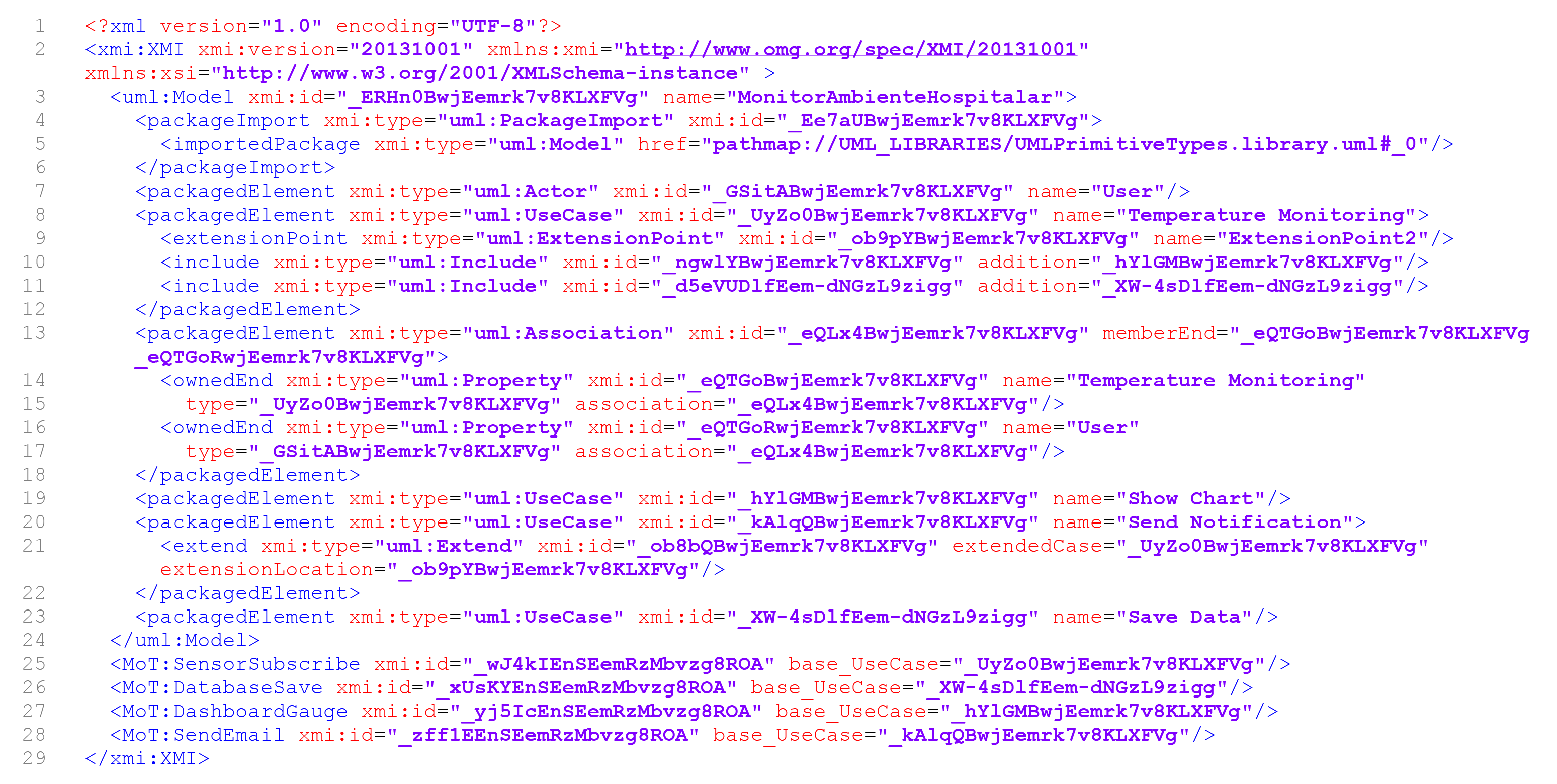}
\caption{XMI representing the elements of the use case diagram.}
\label{fig:ec1-caso-uso-xml}
\end{figure}

Importing the \textit{MoT.Profile} makes it possible to enhance the use-case diagram with semantic annotations provided by the profile. These annotations provide high-level abstractions that describe the application’s functionality within a CoT environment without detailing implementation details. Figure \ref{fig:ec1-caso-uso-estereotipos} presents the annotated use-case diagram, showing the use of the following stereotypes: \textit{SensorSubscribe}, \textit{DatabaseSave}, \textit{DashboardGauge}, and \textit{SendEmail}. These annotations will inform the model transformation processes performed in subsequent steps. Once the use-case diagram is complete, it is exported as an XMI document directly from the UML modeling tool. This file captures all diagram elements in a format that a transformer can interpret. Figure \ref{fig:ec1-caso-uso-xml} shows the exported file, where the diagram elements and applied stereotypes are represented as text. 

\subsubsection{Step 2: Diagram transformation} 

\begin{figure}[h!]
\centering
\includegraphics[width=0.8\textwidth]{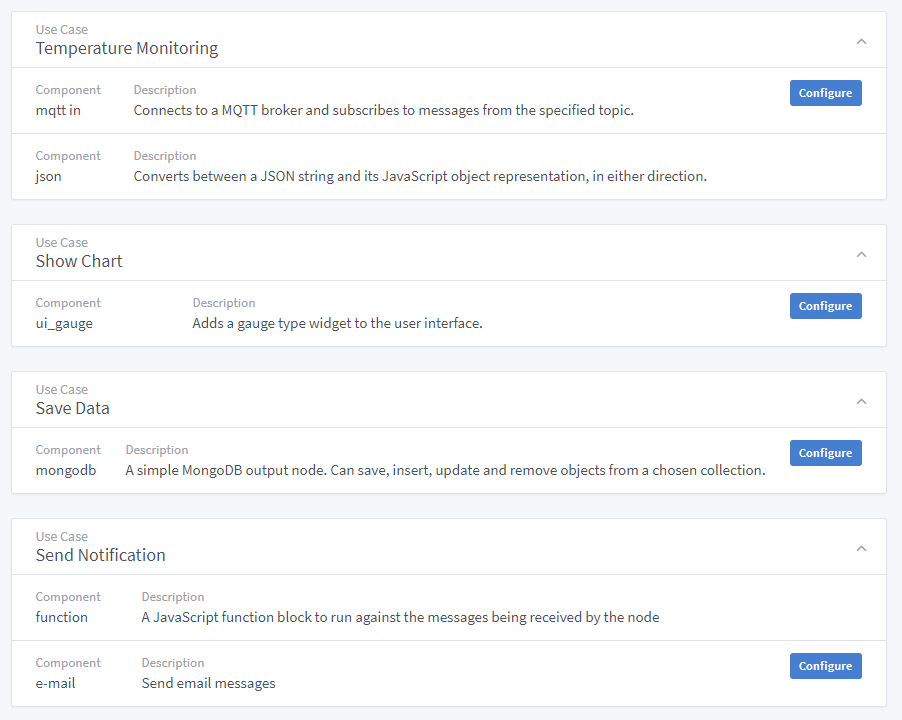}
\caption{List of application components generated after the diagram transformation process, detailing the identified elements and their configurations for deployment.}
\label{fig:ec1-componentes}
\end{figure}

We use the \textit{MoT.Transformer} module to generate the application. By importing the XMI file from the modeling step, the proposed tool performs the first model transformation defined by this approach. In this process, the transformer maps all diagram elements and converts those with UML profile-compatible stereotypes into application components. Figure \ref{fig:ec1-componentes} shows a prototype screen displaying these components, which are transformed from the input XMI file. Figure \ref{fig:ec1-componentes} highlights the diagram's use cases and components created to fulfill the functionality specified by the applied stereotypes.

\subsubsection{Step 3: Component configuration}  

\begin{figure}[h!]
\centering
\includegraphics[width=0.6\textwidth]{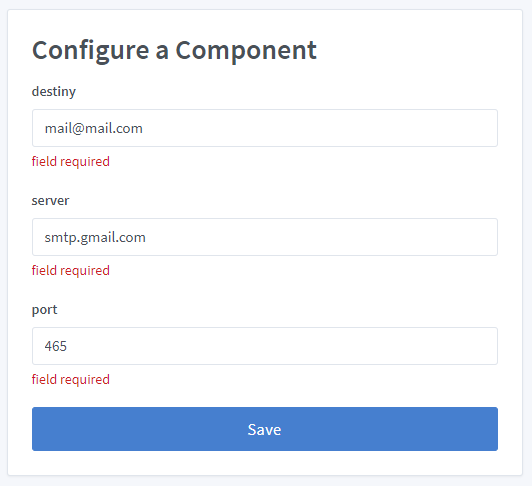}
\caption{User interface for configuring the email component, specifying parameters such as sender details, SMTP settings, and port.}
\label{fig:ec1-email-config}
\end{figure}

After transforming the diagram, the target application includes components required to build each feature defined in the application model. These components vary with the stereotype used and may include functions, cloud services, connectors to other services, and additional components. Not all components are ready for use; some require additional user input. \textit{MoT.Transformer} provides features to help fill in this information. As shown in Figure \ref{fig:ec1-componentes}, components that require configuration feature a button that directs the user to a form requesting the required details. Figure \ref{fig:ec1-email-config} illustrates the information needed for the e-mail component, including the \textit{Send Notification} use case.

As shown in Figure \ref{fig:ec1-email-config}, one key benefit of the MoT approach is abstracting the complexity of specifying programming languages and deployment platforms. With this approach, configuring a resource such as email sending reduces to filling out a simple form, without requiring users to handle technical details. Sensitive information, such as the username and password, has been removed from this stage and must be provided in subsequent steps. Beyond component configuration, another strength of the MoT approach is its ability to abstract the diversity of service configurations across cloud computing providers. By requesting specific information in the form, users can instantiate the services the application needs.

\begin{figure}[h!]
\centering
\includegraphics[width=0.7\textwidth]{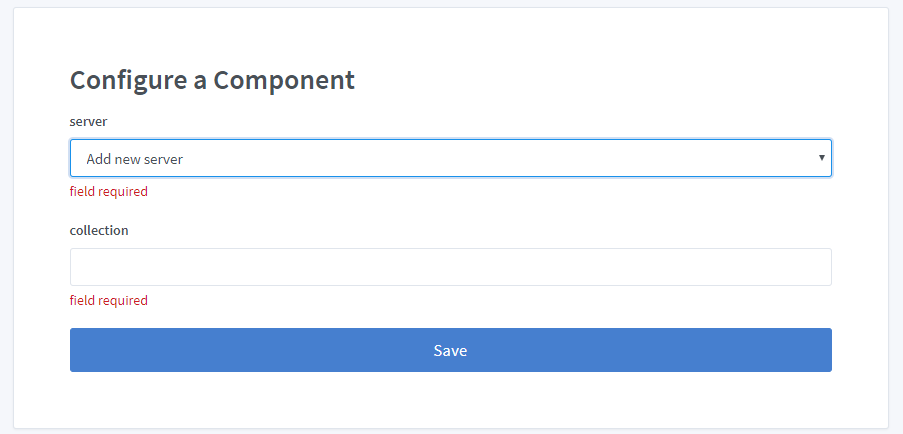}
\caption{Interface for configuring the MongoDB component, specifying server details and target collection for data storage.}
\label{fig:ec1-mongo-config}
\end{figure}

\begin{figure}[h!]
\centering
\includegraphics[width=0.6\textwidth]{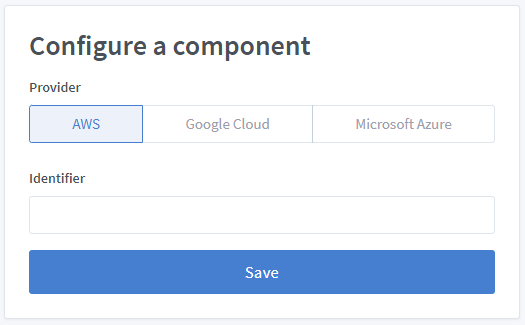}
\caption{Configuring a new MongoDB service instance by selecting the cloud provider and defining a unique identifier.}
\label{fig:ec1-mongo-config2}
\end{figure}

Figure \ref{fig:ec1-mongo-config} shows the configuration of the MongoDB output component, which represents a database instance—a service commonly provided by cloud providers. When configuring this component, users can add a new server. Selecting this option brings up a new form, shown in Figure \ref{fig:ec1-mongo-config2}, where users can choose a cloud service provider to create a new service instance. Once a provider is selected, the \textit{MoT.Transformer} module will configure the service and integrate it with the application component using the connection details.

\textbf{Step 4: Model transformation.} The MoT approach transforms models through the \textit{MoT.Transformer} module. At this stage, it converts the application's components and their configurations into platform-specific components. The approach supports generating applications for the Node-Red platform, including the resources required to run them. In addition to transforming platform-specific components, this step generates the implementation package, which contains infrastructure configuration files for deploying and running the application in a cloud environment. This package includes several tools and frameworks that run on the Node.js platform and are part of the \textit{MoT.Builder} module. It provides the application's file structure, including configuration files, scripts, package dependencies, and more.

\textbf{Step 5: Application deployment.} The \textit{MoT.Builder} module handles the application’s deployment. It includes the Node.js JavaScript runtime and various dependency packages, such as Node-Red and the Serverless Framework, to build the structure. All configuration files and dependencies are included in the implementation package, generated during the model transformation stage. To run the application, run two commands in the operating system's terminal. The first command is \textit{npm run setup}, which installs all necessary dependencies and configures the cloud computing model environment for application execution. The second command is node \textit{./node\_modules/nodered/red.js -s ./settings.js}. This prepares a local environment to execute the configurations. The Node-Red platform will be accessible at \textit{http://localhost:1880}. This environment allows customization of components in the Node-Red application.

\begin{figure}[!h]
\centering
\includegraphics[width=0.9\textwidth]{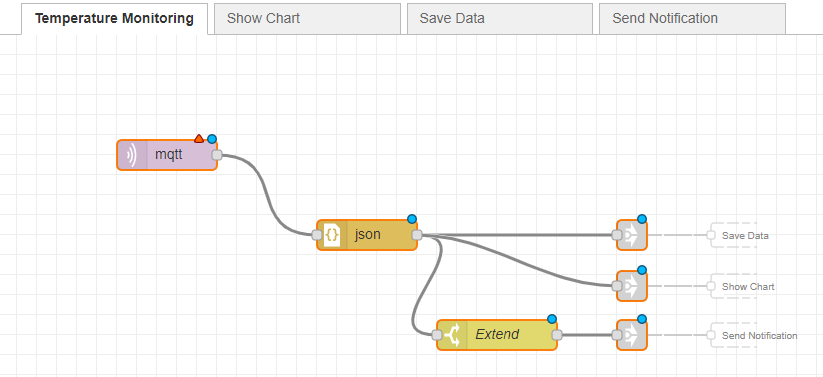}
\caption{Visual representation of a temperature monitoring workflow implemented in the Node-Red platform.}
\label{fig:node-red-interface-temperature}
\end{figure}

\begin{figure}[h!]
\centering
\subfigure[ref1][Show Chart]{\includegraphics[width=0.8\textwidth]{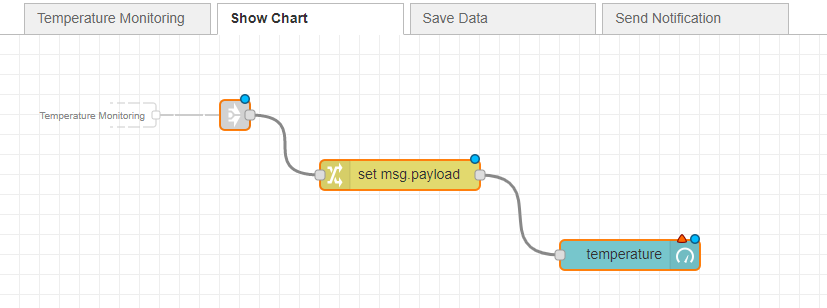}}
\subfigure[ref2][Save Data]{\includegraphics[width=0.8\textwidth]{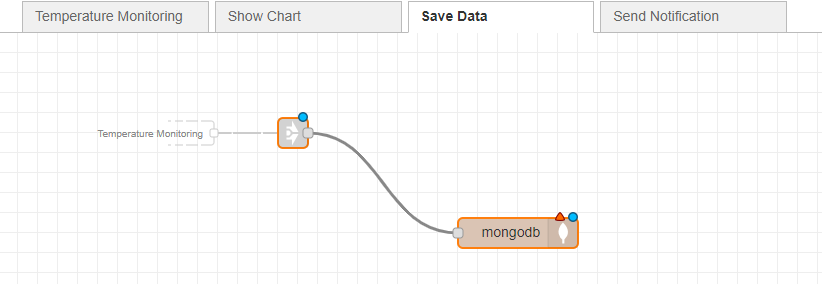}}
\subfigure[ref3][Send Notification]{\includegraphics[width=0.8\textwidth]{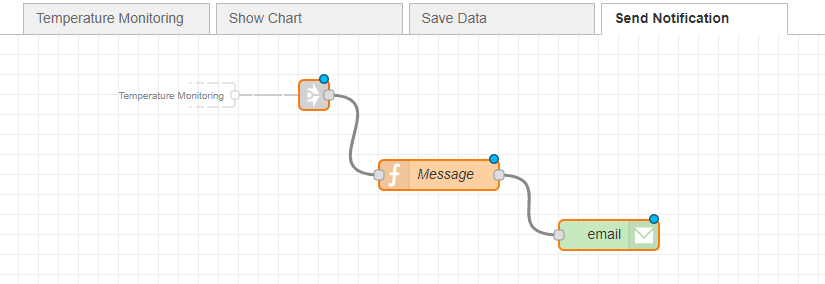}}
\caption{Flows automatically generated in Node-RED for key functionalities: (a) Show Chart, (b) Save Data, and (c) Send Notification.}
\label{fig:fluxos-nodered}
\end{figure}

\textbf{Step 6: Component customization.} Component customization is done through the Node-Red interface, accessible via the localhost address. Figure \ref{fig:node-red-interface-temperature} shows the Node-Red interface, where each use case from the diagram is mapped to a specific tab, and the component flow represents the functionality of the stereotype assigned to the use case. The figure for the Temperature Monitor use case shows the configured elements and their connections with other flows. Figure \ref{fig:fluxos-nodered} details each use case and the components corresponding to each functionality.

Component customization is performed through the Node-Red interface, accessible at localhost. Figure \ref{fig:node-red-interface-temperature} shows the Node-Red interface, where each use case in the diagram is mapped to a specific tab, and the component flow represents the functionality of the stereotype attributed to the use case. The figure shows the Temperature Monitor use case, the configured elements, and the elements representing connections to the other fluxes. Figure \ref{fig:fluxos-nodered} shows the details corresponding to each use case and the components corresponding to each functionality.

\begin{figure}[h!]
\centering
\subfigure[ref1][SMTP Config]{\includegraphics[width=0.45\textwidth]{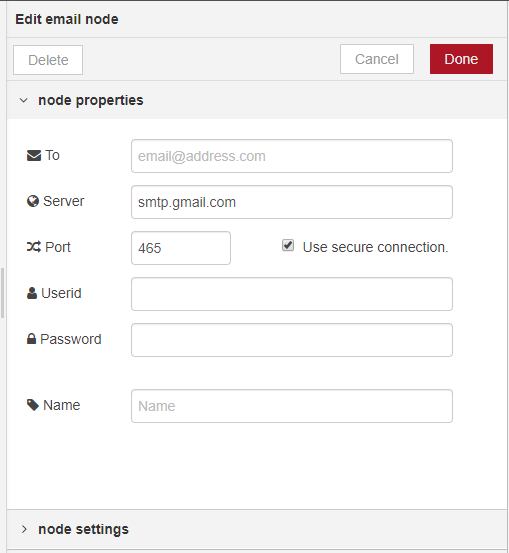}}
\subfigure[ref2][Extend Use Case]{\includegraphics[width=0.45\textwidth]{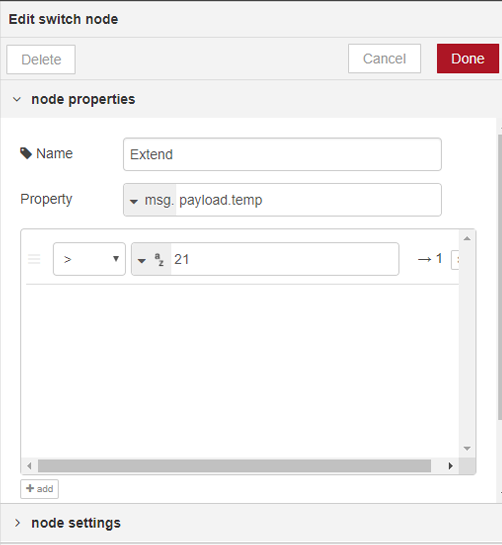}}
\caption{Customization options for the e-mail component: (a) SMTP Configuration and (b) Use Case Extension.}
\label{fig:nodered-config}
\end{figure}

As noted in the component configuration step, some sensitive information, such as login credentials, was not configured earlier. Adjustments must be made at this stage for components requiring this data. Figure \ref{fig:nodered-config}(a) shows the standard Node-RED form, where users can fill in the previously missing fields. Figure \ref{fig:nodered-config}(b) displays the values used to configure the method extension clause for sending a notification when the temperature changes. At this stage, in addition to customizing the components generated by the approach, Node-Red provides other components that can be added to the application flow to create new functionalities not foreseen by the approach. Once the component customization is complete, the application is ready for execution. This marks the final step of the approach: executing the application.

\begin{figure}[h!]
\centering
\includegraphics[scale=0.3]{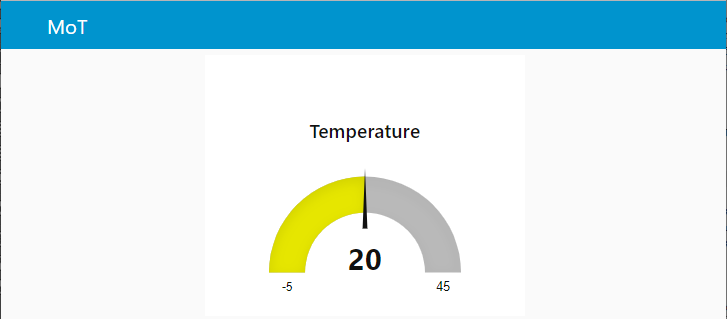}
\caption{Visualization of recorded temperature data over time.}
\label{fig:nodered-dashboard}
\end{figure}

\begin{figure}[h!]
\centering
\includegraphics[width=0.8\textwidth]{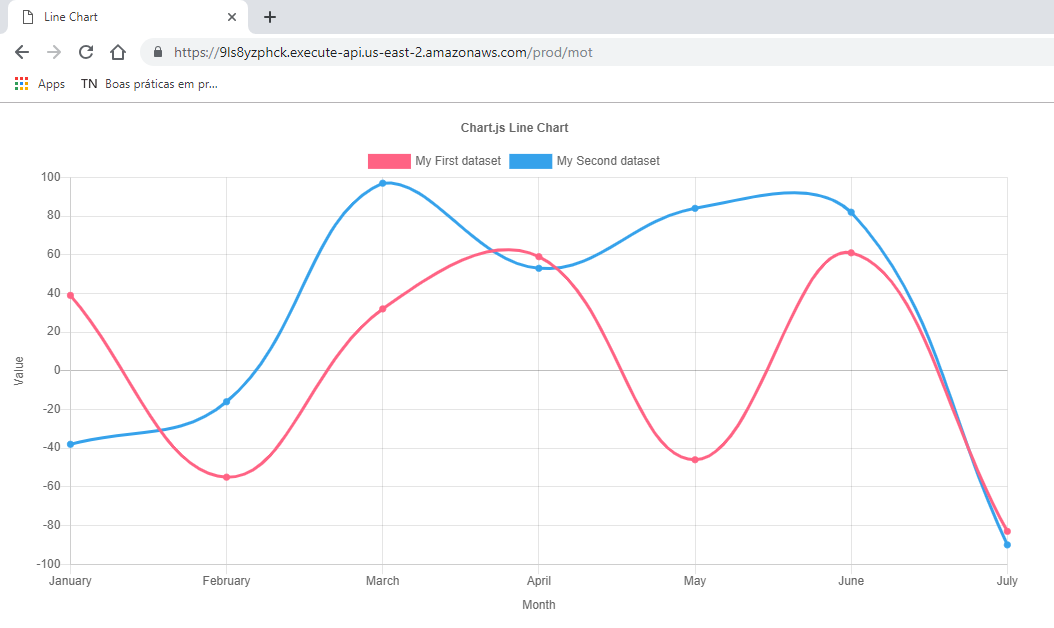}
\caption{Visualization of a static chart application running on the Node-Red platform in a cloud environment.}
\label{fig:nodered-serverlesss-ok}
\end{figure}

\textbf{Step 7: Application execution.} The final step involves executing the application generated through the established process. Execution begins by pressing the \textit{Deploy} button in Node-Red, which starts the local application and updates the environment files set up during the implementation stage. Figure \ref{fig:nodered-dashboard} and Figure \ref{fig:nodered-serverlesss-ok} show the final result after deploying the application in Node-Red. They display an ambient temperature indicator based on the sensor data. In addition to the indicator, when the temperature changes according to the flow extension rule, an email is sent with the details provided in the configuration step.

\subsection{Acceptance of the proposed approach}
\label{subsec_TAM}

We evaluated the MoT approach in an empirical study of 12 participants to assess its acceptance. This evaluation used the Technology Acceptance Model (TAM) \cite{marangunic2015technology} as the framework. The article employed methodologies previously validated in published research to assess technology acceptance, utilizing the TAM questionnaire as a standardized evaluation tool \cite{d2020sw,vieira2025tool,farias2019uml2merge}. Through the TAM questionnaire, we measured perceived ease of use, perceived usefulness, attitude toward use, and behavioral intention to use the MoT approach. Figure \ref{fig_experimental_process} illustrates the experimental process. The phases of the study are outlined as follows:

\begin{figure}[!h]
\centering
\includegraphics[scale=0.27]{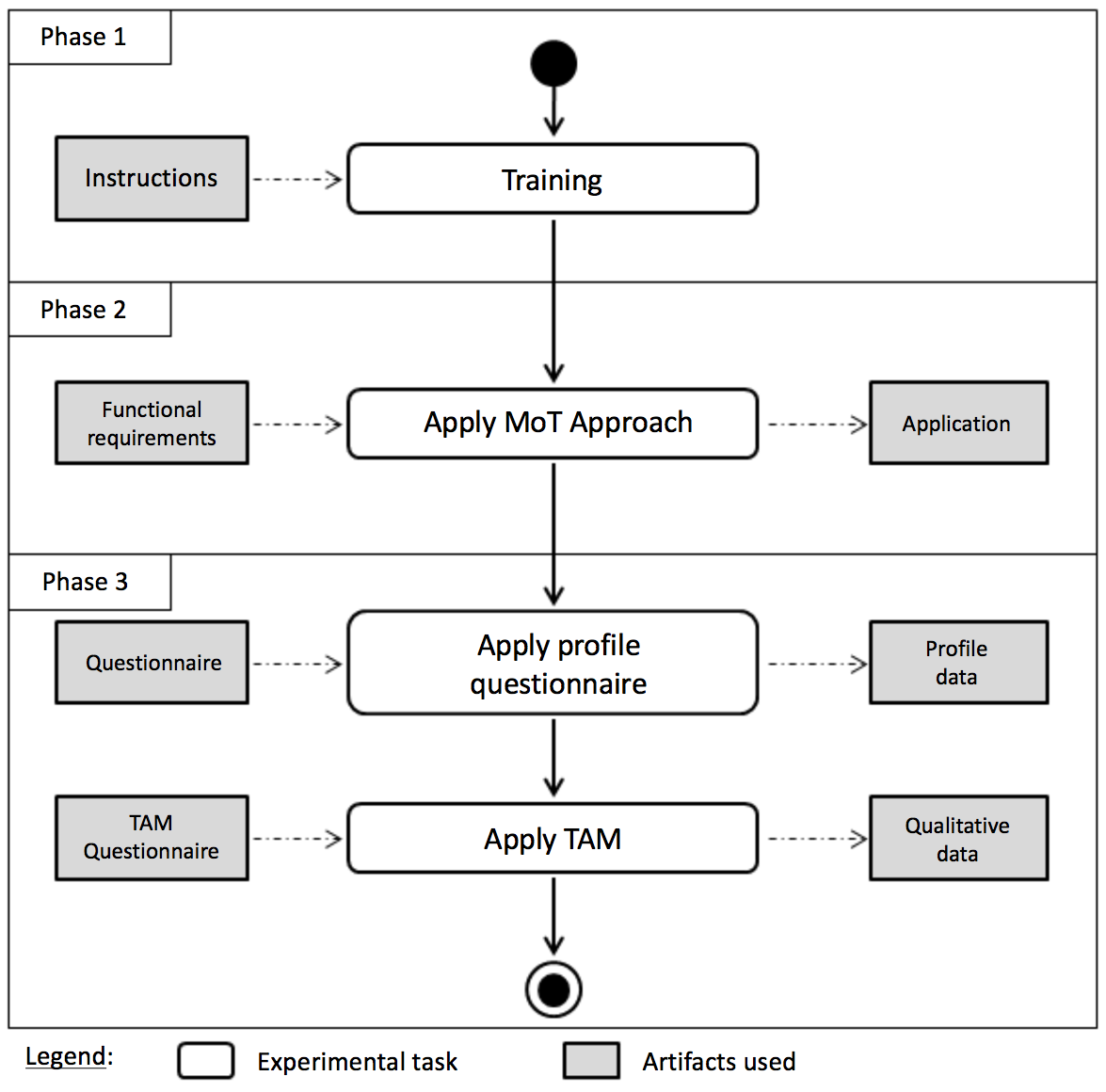}
\caption{Illustration of the experimental process adopted to evaluate the proposed approach, detailing key phases from setup to analysis.}
\label{fig_experimental_process}
\end{figure}

\begin{enumerate}

\item \textbf{Phase 1: Training.} This phase focuses on explaining the experimental study and training participants with the MoT approach. Participants must know their role in the experiment and what is expected of them. At this stage, the authors carefully explained the MoT approach in detail and how to use the \textit{MoT.Transformer} and \textit{MoT.Builder} tools. 

\item \textbf{Phase 2: Apply MoT approach.} Each participant applied the MoT approach to implement the assessment scenario presented in Section~\ref{subsubsec_evalution_scenarios}. Participants received the application's functional requirements (input data) and were guided through the MoT development process, including all steps required to implement the application. At the end of this phase, participants are expected to have a complete understanding of the approach and, in doing so, can assess MoT in the next phase. We understand that participants need to use the approach to assess its usefulness.

\item \textbf{Phase 3: Apply questionnaire.} This phase has two activities. The first activity collects participants' demographic data. Participants answered a set of questions (input) to characterize them. The collected demographic data  (output) is the result of this activity. The second activity uses the technology acceptance model (TAM) as input data. Participants receive a list of questions about perceptions of usability and utility, as well as behavioral intentions related to the prototype. Qualitative data (data output) are generated from users' perspectives on the prototype's usability and acceptance.

\end{enumerate}

\subsubsection{Questionnaires}

We administered two questionnaires: one to characterize participants and another to assess the MoT approach using the Technology Acceptance Model (TAM).

\textbf{Questionnaire 1: Participant profile.} Table \ref{tab:participant_experience} presents the questionnaire used to collect data regarding the participant profile. Analyzing this data is crucial for selecting appropriate candidates for MoT testing. The questions cover general demographics, including age, gender, experience, and educational background, as well as familiarity with IoT technologies, cloud computing, and relevant professional experience. Constructing accurate user profiles ensures that the technique is evaluated by individuals whose backgrounds align with those of potential MoT users.

\textbf{Questionnaire 2: Technology acceptance model.} Table \ref{tab:mot-tam} presents the questionnaire used to assess the usability and acceptability of the approach, based on the Technology Acceptance Model (TAM) proposed by Davis \cite{davis1989perceived} and reviewed by Marangunic and Granic \cite{marangunic2015technology}. The TAM identifies two key factors influencing technology acceptance: (a) perceived ease of use, which reflects the belief that the technology reduces effort, and (b) perceived usefulness, which refers to the belief that the technology enhances performance in task execution.

\subsubsection{Evaluation Scenarios}
\label{subsubsec_evalution_scenarios}

The study participants applied the proposed approach across four evaluation scenarios, allowing them to use and access all its functionalities before completing the TAM evaluation questionnaire.

\begin{itemize}

    \item \textbf{Scenario 01: Application modeling.} This scenario involves extending the traditional use case diagram by incorporating stereotypes from the MoT UML profile using a standard UML modeling tool. The user adds the appropriate stereotypes to each feature in the use case diagram.

    \item \textbf{Scenario 02: Diagram transformation and component configuration.} In this scenario, the user utilizes the \textit{MoT.Transformer} prototype to import the XMI file from the use case diagram and configure the application components. The user imports the XMI representation and completes the component configurations via the respective configuration forms.

    \item \textbf{Scenario 03: Model transformation and application deployment.} Users employ the \textit{MoT.Builder} tool to implement the application and construct its CoT. The user exports the deployment package into the tool and executes commands to generate scripts that create the environments required for the application’s execution.

    \item \textbf{Scenario 04: Customization of components.} This scenario presents greater complexity because it involves specific technical knowledge of the platform to customize the application's components. The user is prompted to customize the Extend component to assign a value to direct the application flow to the new flow.

\end{itemize}

\begin{table}[!h]
\centering
\footnotesize
\begin{tabular}{|l|c|l|}
\hline
\textbf{Category} & \textbf{Number} & \textbf{Question} \\ 
\hline
\multirow{3}{*}{Characterization} 
& 1 & What is your age? \\ \cline{2-3}
& 2 & What is your profession? \\ \cline{2-3}
& 3 & What is your education level? \\ 
\hline
\multirow{4}{*}{Experience} 
& 4 & What is your experience with systems analysis? \\ \cline{2-3}
& 5 & What is your experience with software development? \\ \cline{2-3}
& 6 & What is your experience with the Internet of Things (IoT)? \\\cline{2-3}
& 7 & What is your experience with Cloud Computing? \\ 
\hline
\end{tabular}
\caption{Questionnaire used to characterize participant profiles and experience levels.}
\label{tab:participant_experience}
\end{table}

\begin{table}[ht]
\footnotesize
\centering
\begin{tabular}{|l|c|p{8cm}|}
\hline
\textbf{Category} & \textbf{Number} & \textbf{Question} \\
\hline
\multirow{5}{*}{System Usefulness} 
& 8  & The MoT approach would facilitate the development of CoT applications. \\ \cline{2-3}
& 9  & The MoT approach would help improve developer productivity. \\ \cline{2-3}
& 10 & The MoT approach would reduce the time required to deliver a CoT application. \\ \cline{2-3}
& 11 & The abstraction of IoT components reduces development complexity. \\ \cline{2-3}
& 12 & The abstraction of Cloud Computing components reduces the complexity of configuring environments. \\
\hline

\multirow{3}{*}{Ease of Use} 
& 13 & The MoT prototype is easy to use. \\ \cline{2-3}
& 14 & The MoT approach is easy to learn. \\ \cline{2-3}
& 15 & The MoT approach is easy to master. \\
\hline

\multirow{3}{*}{Overall Perception} 
& 16 & The MoT approach could be used for CoT application development in the software industry. \\ \cline{2-3}
& 17 & A person with little technical knowledge in IoT and Cloud Computing could build a domestic CoT application using MoT. \\ \cline{2-3}
& 18 & I would use the MoT approach to build my CoT application. \\
\hline
\end{tabular}
\caption{Questionnaire for assessing perceived usefulness, ease of use, and overall perception of MoT.}
\label{tab:mot-tam}
\end{table}

\subsubsection{Selection of Participants}

\begin{figure}[h!]
\centering
\includegraphics[scale=0.28]{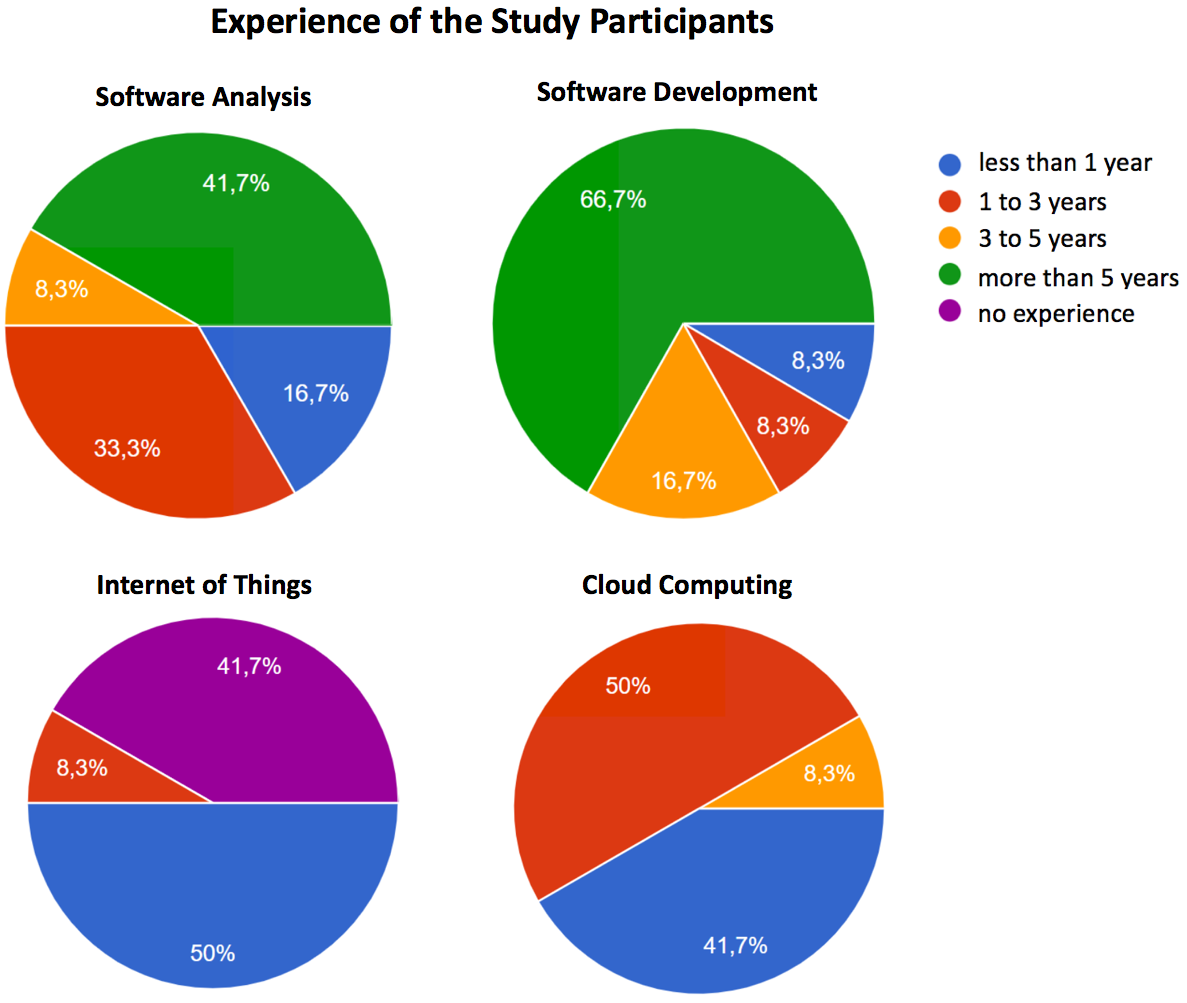}
\caption{Overview of the study participants' experience on a year scale in software analysis, software development, internet of things, and cloud computing.}
\label{fig_participant_profile}
\end{figure}

Participants were selected using convenience sampling from university students with a background in information technology and technology professionals. All participants had experience in software development. Thus, participants with diverse knowledge and experience profiles could be obtained. Figure~\ref{fig_participant_profile} presents an overview of the study participants.

A total of 12 participants took part in the MoT evaluation process. Of these, 10 (83.3\%) were male, and 2 (16.7\%) were female, providing a diverse gender representation. Participants’ ages ranged from 25 to 40 years, reflecting a balanced mix of experience levels. All participants held at least a bachelor's degree in Technology-related fields, ensuring a strong academic foundation. Professionally, 11 participants (91.7\%) worked at technology companies across various sectors, while 1 participant (8.3\%) was a student, providing a fresh perspective for the evaluation.

The age distribution of the participants was as follows: 6 individuals were aged 25 to 29, 2 were aged 30 to 34, 3 were aged 35 to 39, and 1 was aged 40 to 44. Regarding educational background, the majority—9 participants—held an undergraduate degree in Technology, while 3 had a Master’s degree. Professionally, 11 of the 12 participants were in the technology sector, including 6 programmers/systems developers, 4 systems analysts, and 1 software engineer. The remaining participant was a student, adding valuable diversity to the group.

Regarding professional experience in systems analysis, 5 participants had over 5 years, one had between 3 and 5 years, three had between 1 and 3 years, and 2 had less than 1 year. Regarding software development, 8 participants had more than 5 years of experience, 2 had between 3 and 5 years, 1 had between 1 and 3 years, and 1 had less than 1 year. Regarding IoT expertise, 5 participants had no experience, 6 had less than 1 year of experience, and 1 had between 1 and 3 years of experience. In cloud computing, the experience levels were as follows: five participants had less than 1 year of experience, six had between 1 and 3 years of experience, and one had between 3 and 5 years of experience. This diversity in experience levels across different areas highlights a broad spectrum of expertise within the group.

\subsubsection{Results}

This section presents results on participants' perceptions of ease of use, utility, and behavioral intentions regarding the MoT approach. Table \ref{tab:facilidade-uso} summarizes the data collected on participants' perceived ease of use. Notably, 100\% of interviewees agreed that the prototype is easy to use and requires no significant effort to learn its features. Specifically, 50\% of participants partially agreed, while the remaining 50\% fully agreed. Furthermore, when considering the potential to become an expert user, the majority expressed confidence in their ability to master the approach. In fact, 58.3\% fully agreed, and 41.7\% partially agreed that the MoT approach is easy to master.

\begin{table}[h]
    \footnotesize
    \centering
\begin{tabular}{|l|c|c|c|c|c|}
\hline
    \footnotesize
                  & \textbf{Agree} &                  &  &  & \textbf{Disagree} \\
\textbf{Question} & \textbf{Totally} & \textbf{Agree} & \textbf{Neutral} & \textbf{Disagree} & \textbf{Totally} \\
\hline
The prototype of MoT is easy to use & 6 & 6 & 0 & 0 & 0 \\ \hline
MoT's approach is easy to learn     & 6 & 6 & 0 & 0 & 0 \\ \hline
MoT's approach is easy to master    & 7 & 5 & 0 & 0  & 0 \\ 
\hline
\end{tabular}
\caption{Result regarding the perception of ease of use.}
\label{tab:facilidade-uso}
\end{table}

Regarding perceived utility, Table \ref{tab:utilidade} presents the data collected from participants. All participants agreed that the MoT approach would facilitate the development of CoT applications, with 58.3\% fully agreeing and 41.7\% partially agreeing. When evaluating MoT's impact on developer productivity, 58.3\% of participants fully agreed, while the remaining 41.7\% partially agreed. Regarding the potential reduction in delivery time for CoT applications, 50\% fully agreed, while the other 50\% partially agreed. Regarding the abstraction of IoT components, 50\% of participants fully agreed that it reduces the complexity of CoT application development, while the other 50\% partially agreed. Similarly, 58.3\% of participants partially agreed that abstracting cloud computing components reduces the complexity of environment configurations, while 41.7\% fully agreed. These results highlight a strong overall belief in the utility and efficiency of the MoT approach across different aspects of CoT development.

\begin{table}[h]
    \footnotesize
    \centering 
\begin{tabular}{|l|c|c|c|c|c|}
\hline
    \footnotesize
                  & \textbf{Agree} &                  &  &  & \textbf{Disagree} \\
\textbf{Question} & \textbf{Totally} & \textbf{Agree} & \textbf{Neutral} & \textbf{Disagree} & \textbf{Totally} \\
\hline
The MoT approach would facilitate & 7 & 5 & 0 & 0 & 0 \\ 
the development of CoT applications &  &  &  &  &  \\ \hline
MoT's approach would help developer  & 7 & 5 & 0 & 0 & 0 \\ 
productivity &  &  &  &  &  \\ \hline
MoT approach would reduce the time & 6 & 6 & 0 & 0  & 0 \\ 
to deliver a CoT application &  &  &  &  &  \\ \hline
IoT component abstraction reduces & 6 & 6 & 0 & 0  & 0 \\ 
development complexity &  &  &  &  &  \\ \hline
Cloud computing component abstraction & 7 & 5 & 0 & 0 & 0 \\
reduces the complexity of & & & & & \\
environment configuration & & & & & \\
\hline
\end{tabular}
\caption{Result regarding perceived utility.}   
\label{tab:utilidade}
\end{table}

Table \ref{tab:percepcao-geral} presents participants' general perceptions of the MoT approach. All participants agreed that MoT could support CoT application development in the software industry, with 25\% strongly agreeing and 75\% agreeing. Regarding users with limited technical knowledge of IoT and cloud computing developing a CoT application, 41.7\% strongly agreed, 50\% agreed, and 8.3\% remained neutral. Furthermore, 83.3\% of participants intended to use MoT to develop CoT applications, while 16.7\% remained neutral. Overall, these findings indicate positive perceptions of MoT's industrial applicability, accessibility, and potential adoption among the evaluated participants.

\begin{table}[ht]
    \footnotesize
    \centering
\begin{tabular}{|l|c|c|c|c|c|}
        \hline
        \footnotesize
                  & \textbf{Agree} &                  &  &  & \textbf{Disagree} \\
\textbf{Question} & \textbf{Totally} & \textbf{Agree} & \textbf{Neutral} & \textbf{Disagree} & \textbf{Totally} \\
\hline
The MoT approach could be used & 3  & 9 & 0 & 0 & 0 \\
for CoT application development & & & & & \\
in the software industry  & & & & & \\  \hline
A person with little technical & 5 & 6 & 1 & 0 & 0 \\ 
knowledge in IoT development & & & & & \\
and cloud computing could build & & & & & \\
a home CoT application & & & & & \\  \hline
I would use the MoT approach & 1 & 9 & 2 & 0 & 0 \\ 
to create a CoT application & & & & & \\ 
\hline
\end{tabular}
\caption{Result regarding general perception.}    
\label{tab:percepcao-geral}
\end{table}

Therefore, the collected and analyzed data strongly suggest that the MoT approach has significant potential for acceptance among individuals with profiles similar to the participants'. The results highlight the approach’s usability and its promising real-world application, underscoring its potential for broad adoption and practical implementation.

\section{Conclusion and Future Work}
\label{sec:conclusao}

This study explored model-driven development approaches for creating IoT applications that harness cloud computing capabilities. To achieve this, we proposed a model-driven approach called MoT, specifically designed to address critical challenges in CoT applications, with a particular focus on managing cloud service heterogeneity. Our research aimed to investigate whether a model-driven approach could streamline the development of native cloud IoT applications with minimal manual intervention and whether individuals with limited IoT experience could successfully create applications using this approach. To evaluate MoT's effectiveness, we conducted a case study and applied the Technology Acceptance Model.

Our initial results showed that the MoT approach effectively facilitates creating CoT applications with minimal manual effort, as shown in the case study. However, the approach currently lacks full functionality for running applications directly in the cloud. Despite this limitation, the TAM-based evaluation yielded positive feedback from study participants. While this is an initial assessment, the results indicate that MoT is perceived as both useful and easy to use, even among individuals with limited IoT knowledge. These findings affirm the approach's potential and suggest promising acceptance among technology professionals, offering a favorable response to the second research question.

As part of our ongoing research agenda, future work will expand MoT beyond its current execution-oriented foundation to incorporate an event-driven architectural style \cite{lazzari2023} and broader operational coverage. First, we plan to investigate formal concurrency modeling and synchronization mechanisms to better capture parallel execution behaviors, coordination requirements, and potential contention scenarios that arise in Cloud-of-Things environments. Second, we intend to strengthen systematic support for complex multi-device configurations by addressing distributed orchestration, inter-device dependencies, and scalable deployment strategies. Finally, we will explore architectural strategies to better manage heterogeneity across different cloud backends, including resilience and fault-tolerance mechanisms that handle failures, variable service availability, and diverse platform constraints. Together, these research directions aim to mature MoT toward a more comprehensive and robust framework capable of supporting large-scale, heterogeneous, and mission-critical CoT applications. Researchers can also explore ways to enhance MoT and evaluate its effectiveness across diverse contexts. These include: (1) incorporating additional UML diagrams to broaden the modeling capabilities, (2) comparing MoT with traditional IoT development methods to assess its relative advantages, (3) expanding the diversity of participants to include both general users and experienced IoT professionals, (4) the need for a more flexible and configurable transformation engine, (5) mechanisms to support rollback, resilience, and fault handling during deployment, and (6) the design of systematic performance and scalability evaluations. This article opens a clear path for future efforts to formalize optimization objectives, define measurable constraints, evaluate correctness, and introduce quantitative performance baselines.

These improvements would provide deeper insights into MoT's strengths and limitations, bringing the research closer to establishing a solid foundation for advancing Model-Driven Development in IoT application creation. Rather than constituting a definitive study, our work serves as an initial step toward a more comprehensive research agenda in this area.

\section*{Ethics Issue}
\label{sec:ethics}

We conducted the research in settings where ethics approval for survey studies was deemed unnecessary because the investigation involved voluntary participation and collected no sensitive information.

\section*{Acknowledgement}

This work was supported in part by the Conselho Nacional de Desenvolvimento Cient\'{i}fico e Tecnol\'{o}gico (CNPq), under Grant No. 312320/2025-6.

\end{document}